\documentclass[preprint,12pt]{elsarticle}
\usepackage{amsmath,amssymb,amsfonts}
\usepackage{algorithm}
\usepackage{algorithmic}
\usepackage{graphicx}
\usepackage{textcomp}
\usepackage{xcolor}
\usepackage{tikz}
\usepackage{etoolbox}
\usepackage{tikz-3dplot} 

\usetikzlibrary{
  shapes.geometric, 
  arrows.meta, 
  positioning, 
  fit, 
  backgrounds, 
  calc, 
  patterns,
  bending,
  shadings,
  decorations.markings
}

\tikzset{
  every node/.style={
    font=\ttfamily, 
    text=black,
    align=center,
  },
  every path/.style={
    line width=.8pt,
  }
}

\journal{Aerospace Science and Technology}

\begin{document}

\begin{frontmatter}

\title{Learned Memory Attenuation in Sage-Husa Kalman Filters for Robust UAV State Estimation}            

\author[WUT]{Kenan Majewski}
\ead{kenan.majewski.dokt@pw.edu.pl}

\author[WUT]{Marcin Żugaj}

\affiliation[WUT]{organization={Institute of Aeronautics and Applied Mechanics, Warsaw University of Technology},
            addressline={Nowowiejska 24}, 
            city={Warsaw},
            postcode={00-665}, 
            country={Poland}}

\begin{abstract}
  Unmanned Aerial Vehicles in dynamic environments face telemetry outages, structural vibrations, and regime-dependent noise that invalidate the stationary covariance assumptions of classical Kalman filters. The Sage-Husa Kalman Filter (SHKF) estimates noise statistics online, but its reliance on a static, scalar forgetting factor forces a strict compromise between steady-state stability and transient responsiveness. We introduce the N-Deep Recurrent Sage-Husa Filter (NDR-SHKF), which replaces this scalar parameter with a vector-valued memory attenuation policy learned by a hierarchical recurrent network operating on whitened innovation sequences. A bifurcated architecture routes shallow recurrent states to capture instantaneous sensor anomalies and deep states to encode sustained dynamic trends, while an auxiliary reconstruction objective prevents feature collapse. The complete filter, including recursive covariance updates, is trained end-to-end via backpropagation through time to directly minimize state estimation error. Evaluations on topologically distinct chaotic attractors demonstrate cross-domain generalization, outperforming purely data-driven baselines that diverge under out-of-distribution dynamics. Furthermore, evaluations on recorded real-world UAV flight datasets validate the framework's practical viability, demonstrating its capacity to bridge transitions into proprioceptive dead reckoning and outperform classical adaptive estimators during sensor outages.
\end{abstract}


\end{frontmatter}

\section{Introduction}
\label{sec:introduction}

Accurate state estimation forms the foundation of autonomous flight and robotic navigation~\cite{ireland2012development,wahbah2022real}. For Unmanned Aerial Vehicles (UAVs) navigating complex, dynamically changing environments, robust estimation of position, velocity, and attitude in the presence of sensor noise, hardware faults, and unmodeled dynamics is required~\cite{yang2024adaptive,wang2025adaptive,hajiyev2013robust}. While the Extended Kalman Filter (EKF)~\cite{bar2001estimation} constitutes the industry standard, its mathematical optimality depends on the assumption that the process noise covariance ($\mathbf{Q}$) and measurement noise covariance ($\mathbf{R}$) are strictly known and stationary. In operational aerospace scenarios, this condition is systematically violated. UAVs frequently undergo rapid dynamic regime shifts, alternating between steady-state hovering and high-acceleration maneuvers~\cite{foehn2022agilicious,kaufmann2023champion}. During such transients, aerodynamic disturbances, state-dependent drag, and rotor-induced structural vibrations alter the underlying sensor noise profiles, invalidating the stationary covariance assumptions of classical filters~\cite{brossard2020ai,qin2018vins}.

To mitigate the divergence caused by static noise assumptions, Adaptive Kalman Filtering (AKF) methods~\cite{mehra1972approaches,dunik2017noise}, notably the Sage-Husa Kalman Filter (SHKF)~\cite{sage1969adaptive}, have been developed to estimate drifting noise statistics online. In aerial navigation, the Inertial Measurement Unit (IMU) serves as the primary proprioceptive sensor, providing the high-frequency acceleration and angular rate data essential for flight control. However, commercial-grade MEMS IMUs are frequently affected by time-varying biases, thermo-mechanical noise, and high-frequency vibrations induced by rotor dynamics~\cite{el2008analysis,leishman2014quadrotors}. By recursively refining the process and measurement covariance matrices using the sequence of measurement innovations, the statistical discrepancy between predicted states and actual sensor readings, the SHKF continuously re-calibrates the estimator's confidence bounds~\cite{mohamed1999adaptive} to account for these dynamic IMU errors. This adaptive mechanism has been deployed across diverse domains, including ultra-wideband (UWB) localization in GPS-denied environments~\cite{juston2025robust}, electric vehicle servo control~\cite{wang2022modified}, and robust filter designs for uncertain noise distributions~\cite{fan2024improved}.

Nevertheless, the theoretical efficacy of the SHKF is predominantly bottlenecked by its strict mathematical reliance on a scalar forgetting factor, $b$, which dictates the estimator's effective memory length. This introduces a fundamental trade-off. A conservative factor (approaching unity) ensures stable error covariance estimates during steady flight but induces lag and potential divergence during aggressive maneuvers. Conversely, a smaller factor guarantees rapid responsiveness to dynamic shifts but risks numerical instability and covariance collapse by over-amplifying high-frequency sensor noise. Conventionally, $b$ is tuned statically offline~\cite{abbeel2005discriminative} or modulated via rigid, hand-crafted heuristics such as an adaptive fading factor~\cite{xia1994adaptive}. These rudimentary mechanisms lack the representational capacity to distinguish multi-scale temporal correlations; they cannot differentiate between a transient, high-amplitude sensor anomaly and a sustained physical maneuver, ultimately restricting the autonomous system's operational envelope.

Recent advancements in differentiable programming~\cite{haarnoja2016backprop,jonschkowski2018differentiable,wen2021end} have driven a paradigm shift toward data-driven filtering, aiming to overcome these classical limitations using artificial intelligence~\cite{cohen2024inertial,shlezinger2025artificial}. Purely neural architectures, such as KalmanNet~\cite{revach2022kalmannet,mortada2025recursive} and contemporary Recurrent Neural Network (RNN) trackers~\cite{zheng2019rnn,song2026rg}, have demonstrated that deep networks can effectively capture the complex mapping from innovation sequences directly to state updates or Kalman gains, proving effective when the underlying dynamics are partially unmodeled. However, by circumventing the standard Riccati covariance recursions, these black-box approaches simultaneously forfeit the stability bounds, convergence guarantees, and formal uncertainty quantification inherent to Bayesian filters.

To preserve these properties, intermediate hybrid approaches have sought to couple neural networks with the SHKF. Yet, the current intersection of deep learning and adaptive filtering remains dominated by loosely integrated architectures. Frameworks that couple adaptive Kalman filters with Elman neural networks~\cite{gao2012seam} or standard Feedforward Networks~\cite{kim2025battery} generally treat the classical filter as a fixed preprocessing stage rather than a tightly integrated differentiable component. Similarly, advanced spatio-temporal architectures, such as hybrid CNN-GRU pipelines or Long Short-Term Memory (LSTM) networks combined with AKFs~\cite{meng2024gnss,li2025rov,hosseini2025online}, operate primarily as parallel redundant systems. They utilize neural networks to predict pseudo-measurements during GNSS/INS sensor blackouts, acting as synthetic substitutes rather than intervening in the mathematical core of the filter itself. This creates a persistent dichotomy in the literature: classical filters remain numerically stable but parametrically limited, while neural estimators are adaptive but theoretically opaque.

Consequently, the formulation of a differentiably learned, internal memory attenuation policy remains underexplored. Rather than replacing the filter or operating alongside it, the optimal selection of the adaptation factor must be treated as a sequential decision-making problem. Viewed through the lens of meta-learning~\cite{hospedales2021meta,finn2017model}, a neural architecture can be trained not to estimate the state directly, but to dynamically govern the recursive mathematics of the SHKF, achieving simultaneous resilience and numerical stability.

In this work, we introduce the N-Deep Recurrent Sage-Husa Filter (NDR-SHKF), a novel framework designed to dynamically modulate the memory of an adaptive filter through learned representations. Rather than supplanting the robust recursive mathematics of the SHKF with a purely opaque neural covariance generator, the proposed architecture employs a hierarchical, N-deep RNN functioning strictly as a hyper-controller for the adaptation rate.

Our approach introduces three primary contributions to the field of adaptive estimation:
\begin{enumerate}
    \item \textbf{Whitened Multi-Scale Inputs:} The methodology employs the Cholesky decomposition of the innovation covariance to normalize the system inputs, providing the neural network with a scale-invariant whitened innovation vector. This mathematical transformation enables the architecture to robustly generalize across diverse sensor modalities and disparate noise magnitudes.
    \item \textbf{N-Deep Temporal Context:} Through the implementation of stacked Gated Recurrent Unit (GRU)~\cite{cho2014learning} layers, the network disentangles dynamics across multiple timescales. Shallow layers are optimized to capture immediate sensor anomalies, whereas deeper layers are responsible for identifying long-term maneuvering trends, facilitating context-aware memory attenuation.
    \item \textbf{End-to-End Meta-Learning:} Drawing upon meta-learning frameworks~\cite{finn2017model,wang2016learning}, the complete architecture, encompassing the RNN policy and the recursive Sage-Husa update equations, is formulated as a unified differentiable computational graph. This structural integration permits the policy to be optimized via analytical gradients, directly minimizing state estimation error while maintaining the numerical stability of the recursive covariance updates.
\end{enumerate}%

The remainder of this paper is organized as follows: Section~\ref{sec:theoretical} provides the theoretical framework, detailing the EKF and conventional Sage-Husa algorithms. Section~\ref{sec:ndrshkf} describes the proposed NDR-SHKF architecture, the end-to-end differentiable optimization strategy, and an analysis of its computational complexity. Section~\ref{sec:sim_experiments} presents a cross-domain generalization analysis, evaluating out-of-distribution robustness across distinct chaotic dynamical systems. Section~\ref{sec:uav_experiments} evaluates the proposed framework using recorded flight telemetry from a physical UAV, demonstrating its capacity to maintain robust state estimation and transition to proprioceptive dead reckoning during external telemetry loss. Finally, Section~\ref{sec:conclusions} summarizes the findings.

\section{Theoretical Framework}\label{sec:theoretical}

\subsection{Problem Formulation}\label{subsec:problem_formulation}
Consider a discrete-time nonlinear dynamic system described by the state-space model:
\begin{align}
    \mathbf{x}_{k} &= f(\mathbf{x}_{k-1}) + \mathbf{w}_{k-1}, \\
    \mathbf{z}_{k} &= h(\mathbf{x}_k) + \mathbf{v}_k,
\end{align}
where $\mathbf{x}_k \in \mathbb{R}^{n_x}$ is the state vector and $\mathbf{z}_k \in \mathbb{R}^{n_z}$ is the measurement vector at time step $k$. The nonlinear mappings $f : \mathbb{R}^{n_x}\rightarrow\mathbb{R}^{n_x}$ and $h : \mathbb{R}^{n_x}\rightarrow\mathbb{R}^{n_z}$ denote the state transition and observation functions, respectively.

In standard optimal estimation theory, the process noise $\mathbf{w}_k$ and measurement noise $\mathbf{v}_k$ are assumed to be mutually uncorrelated, zero-mean Gaussian white noise sequences with strictly known, stationary covariance matrices $\mathbf{Q}$ and $\mathbf{R}$:
\begin{equation}
    \mathbb{E}[\mathbf{w}_k\mathbf{w}_j^\top] = \mathbf{Q} \delta_{kj}, \quad \mathbb{E}[\mathbf{v}_k\mathbf{v}_j^\top] = \mathbf{R} \delta_{kj},
\end{equation}
where $\delta_{kj}$ is the Kronecker delta. 

In dynamic tracking scenarios, these stationary Gaussian assumptions are frequently invalidated. Unmodeled dynamics, transient sensor anomalies, and state-dependent disturbances render the true noise covariances, $\mathbf{Q}_k$ and $\mathbf{R}_k$, unknown and non-stationary. Consequently, the objective is to formulate an adaptive recursive estimator that jointly infers the system state and these dynamic noise statistics directly from the innovation sequence, without relying on domain-specific static priors.

\subsection{Extended Kalman Filter (EKF)}\label{subsec:ekf}
The Extended Kalman Filter (EKF) establishes the foundational recursive framework for addressing system nonlinearities via first-order Taylor series linearization. The Jacobians of the system models, evaluated at the most recent state estimates, are:
\begin{equation}
    \mathbf{F}_{k-1} = \left.\frac{\partial f}{\partial \mathbf{x}}\right|_{\mathbf{x} = \hat{\mathbf{x}}_{k-1|k-1}}, \quad \mathbf{H}_k = \left.\frac{\partial h}{\partial \mathbf{x}}\right|_{\mathbf{x} = \hat{\mathbf{x}}_{k|k-1}}.
\end{equation}

The recursive algorithm consists of a prediction phase and an update phase. The \textit{a priori} state estimate and error covariance $\mathbf{P}$ are projected forward in time:
\begin{align}
    \hat{\mathbf{x}}_{k|k-1} &= f(\hat{\mathbf{x}}_{k-1|k-1}) \label{eq:ekf_predict_state},\\
    \mathbf{P}_{k|k-1} &= \mathbf{F}_{k-1} \mathbf{P}_{k-1|k-1} \mathbf{F}_{k-1}^\top + \mathbf{Q}_{k-1} \label{eq:ekf_predict_cov}.
\end{align}

Upon acquiring a new measurement $\mathbf{z}_k$, the filter computes the innovation sequence $\boldsymbol{\nu}_k$ and its associated predicted covariance $\mathbf{S}_k$:
\begin{align}
    \boldsymbol{\nu}_k &= \mathbf{z}_k - h(\hat{\mathbf{x}}_{k|k-1}), \\
    \mathbf{S}_k &= \mathbf{H}_k \mathbf{P}_{k|k-1} \mathbf{H}_k^\top + \mathbf{R}_k .
\end{align}

The optimal Kalman gain $\mathbf{K}_k \in \mathbb{R}^{n_x \times n_z}$, which minimizes the trace of the \textit{a posteriori} error covariance, the state update and covariance update, are evaluated as:
\begin{align}
    \mathbf{K}_k &= \mathbf{P}_{k|k-1} \mathbf{H}_k^\top \mathbf{S}_k^{-1}, \\
    \hat{\mathbf{x}}_{k|k} &= \hat{\mathbf{x}}_{k|k-1} + \mathbf{K}_k \boldsymbol{\nu}_k, \\
    \mathbf{P}_{k|k} &= (\mathbf{I} - \mathbf{K}_k \mathbf{H}_k) \mathbf{P}_{k|k-1}.
\end{align}

Because the mathematical optimality of the standard EKF depends on accurate, \textit{a priori} knowledge of $\mathbf{Q}_k$ and $\mathbf{R}_k$, static formulations inevitably degrade or diverge under the unknown time-varying conditions.

\subsection{Sage-Husa Kalman Filter (SHKF)}\label{subsec:shkf}
To address the necessity for online covariance estimation, the Sage-Husa Kalman Filter (SHKF) recursively approximates the unknown noise moments via a fading memory approach, modulated by an adaptation factor $d_k$:
\begin{align}
    \mathbf{R}_{k} &= (1 - d_k)\mathbf{R}_{k-1} + d_k \hat{\mathbf{R}}_{k},\\
    \mathbf{Q}_{k} &= (1 - d_k)\mathbf{Q}_{k-1} + d_k \hat{\mathbf{Q}}_{k},
\end{align}
where the instantaneous empirical covariance estimates are derived from the filter's internal state components:
\begin{align}
    \hat{\mathbf{R}}_{k} &= \boldsymbol{\nu}_k\boldsymbol{\nu}_k^\top - \mathbf{H}_k\mathbf{P}_{k|k-1}\mathbf{H}_k^\top ,\\
    \hat{\mathbf{Q}}_{k} &= \mathbf{K}_k\boldsymbol{\nu}_k\boldsymbol{\nu}_k^\top\mathbf{K}_k^\top + \mathbf{P}_{k|k} - \mathbf{F}_{k-1}\mathbf{P}_{k-1|k-1}\mathbf{F}_{k-1}^\top.
\end{align}

In classical formulations, the adaptation parameter is defined explicitly as 
\begin{equation}
    d_k = (1-b)/(1-b^{k+1}),
\end{equation}
converging to $1-b$ as $k \to \infty$, where $b \in (0, 1)$ acts as a static, scalar forgetting factor.

As previously established in Section~\ref{sec:introduction}, this rigid scalar parameterization imposes an unavoidable compromise between steady-state noise suppression and transient responsiveness. Beyond this structural limitation, the conventional SHKF suffers from inherent numerical instability in real-world applications. Because the empirical covariance updates require subtracting the predicted \textit{a priori} uncertainty from the instantaneous innovation outer product, finite-sample statistical fluctuations and linearization errors frequently yield indefinite or negative-definite matrices~\cite{fan2024improved,hosseini2025online}. To prevent filter divergence, robust filter designs must rely on heuristic algebraic safeguards, such as explicit magnitude clipping, enforcing minimum variance thresholds on the diagonal, and periodic re-symmetrization, to guarantee positive definiteness prior to subsequent matrix inversions. In this work, these safeguards are implemented by maintaining $\mathbf{Q}_k$ and $\mathbf{R}_k$ as diagonal matrices, with their diagonal elements strictly bounded below by a positive floor parameter and constrained within a factor of their nominal base values.

\section{N-Deep Recurrent Sage-Husa Kalman Filter (NDR-SHKF)}
\label{sec:ndrshkf}

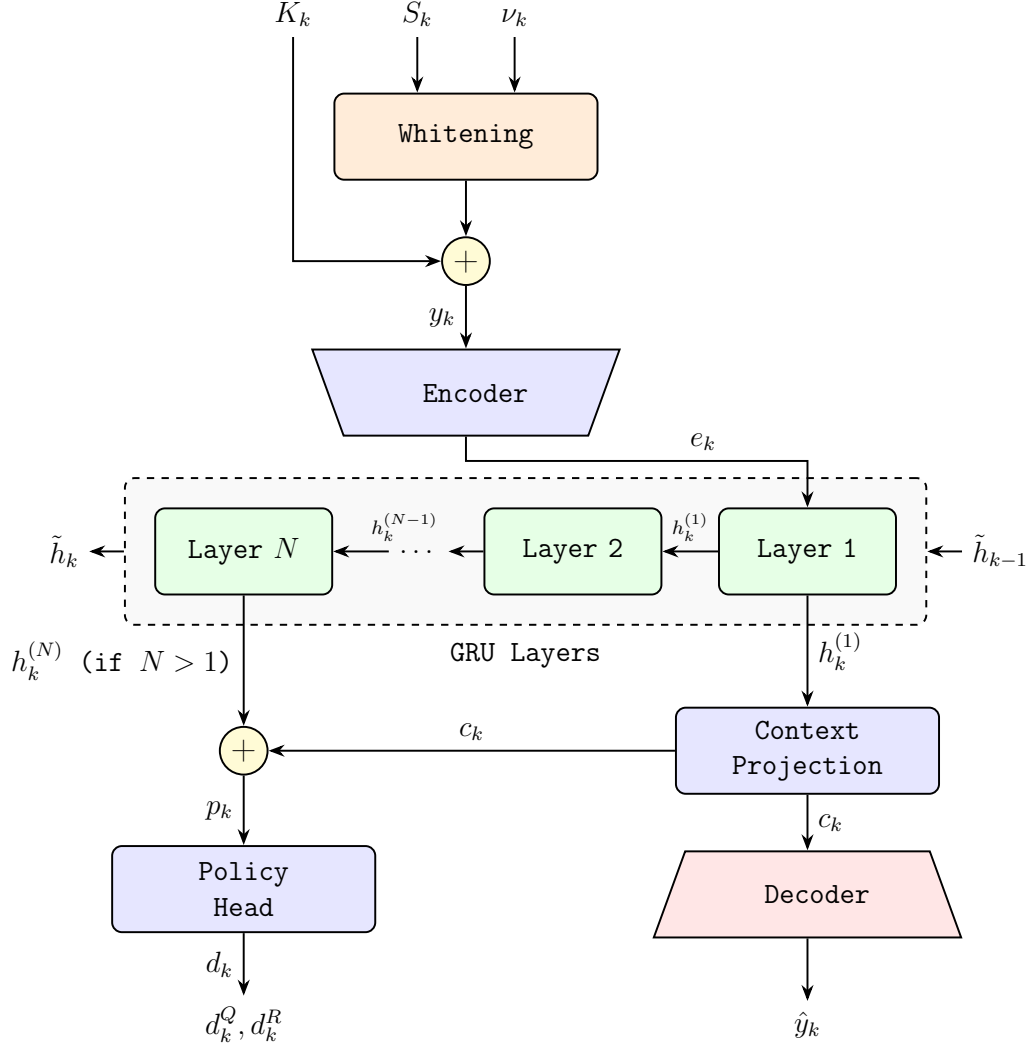
\begin{figure}[t!]
    \centering
    \resizebox{1.0\columnwidth}{!}{%
        \begin{tikzpicture}[
            >=Stealth,
            node distance=1.5cm and 2cm,
            block/.style={rectangle, draw, fill=blue!10, text width=8.5em, text centered, rounded corners, minimum height=3em, thick, font=\ttfamily},
            whiteblock/.style={block, fill=orange!15},
            encblock/.style={trapezium, trapezium left angle=110, trapezium right angle=110, draw, fill=blue!10, text width=3em, text centered, minimum height=3em, thick, font=\ttfamily},
            decblock/.style={trapezium, trapezium left angle=70, trapezium right angle=70, draw, fill=red!10, text width=3em, text centered, minimum height=3em, thick, font=\ttfamily},
            grublock/.style={rectangle, draw, fill=green!10, text width=5.5em, text centered, rounded corners, minimum height=3em, thick, font=\ttfamily},
            concat/.style={circle, draw, fill=yellow!20, text centered, thick, inner sep=2pt, font=\bfseries},
            io/.style={text centered, font=\ttfamily},
            line/.style={draw, ->, thick}
        ]

        \node[whiteblock] (prep) at (0,0) {Whitening};
        \node[io, above=0.8cm of prep, xshift=-0.7cm] (S) {$S_k$};
        \node[io, above=0.8cm of prep, xshift=0.7cm] (innov) {$\nu_k$};
        \node[io, left=1.0cm of S] (K) {$K_k$};

        \node[concat, below=0.8cm of prep] (concat1) {+};

        \draw[line] (innov) -- ($(prep.north)+(0.7,0)$);
        \draw[line] (S) -- ($(prep.north)+(-0.7,0)$);
        \draw[line] (prep.south) -- (concat1.north);
        \draw[line] (K.south) |- (concat1.west);

        \node[encblock, below=0.9cm of concat1] (enc) {Encoder};
        \draw[line] (concat1.south) -- node[left] {$y_k$} (enc.north);

        \node[below=1.5cm of enc] (grucenter) {};
        \node[grublock, right=0.1cm of grucenter] (gru1) {Layer 2};
        \node[grublock, right=0.8cm of gru1] (gru0) {Layer 1};
        \node[left=0.1cm of grucenter, font=\bfseries] (dots) {$\dots$};
        \node[grublock, left=0.8cm of dots] (gruN) {Layer $N$};

        \begin{scope}[on background layer]
            \node[draw, thick, dashed, rounded corners, fill=gray!5, inner sep=12pt, fit=(gru0) (gruN)] (gru) {};
        \end{scope}
        \node[below=0.1cm of gru.south, font=\ttfamily\bfseries] {GRU Layers};

        \node[io, right=0.5cm of gru] (hin) {$\tilde{h}_{k-1}$}; 
        \node[io, left=0.5cm of gru] (hout) {$\tilde{h}_k$};

        \draw[line] (gru0.west) -- node[above, font=\scriptsize] {$h^{(1)}_k$} (gru1.east);
        \draw[line] (gru1.west) -- (dots.east);
        \draw[line] (dots.west) -- node[above right, font=\scriptsize] {$h^{(N-1)}_k$} (gruN.east);

        \draw[line] (enc.south) -- ++(0,-0.35)  -| node[above,xshift=-1.5cm] {$e_k$} (gru0.north);

        \draw[line] (hin) -- (gru);
        \draw[line] (gru) -- (hout);


        \node[block, below=1.6cm of gru0] (ctx) {Context\\Projection};
        \node[decblock, below=0.8cm of ctx] (dec) {Decoder};
        \node[io, below=0.9cm of dec] (rec) {$\hat{y}_k$};

        \node[concat] (concat2) at (gruN |- ctx) {+}; 
        \node[block, below=1.0cm of concat2] (dk) {Policy\\Head};
        \node[io, below=0.9cm of dk] (outdk) {$d_k^Q, d_k^R$};


        \draw[line] (gru0.south) -- node[right] {$h^{(1)}_k$} (ctx.north);
        \draw[line] (gruN.south) -- node[left] {$h_{k}^{(N)}$ (if $N>1$)} (concat2.north);

        \draw[line] (ctx.south) -- node[right] {$c_k$} (dec.north);
        \draw[line] (dec.south) -- (rec.north);

        \draw[line] (ctx.west) -- node[above] {$c_k$} (concat2.east);

        \draw[line] (concat2.south) -- node[left] {$p_k$} (dk.north);
        \draw[line] (dk.south) -- node[left] {$d_k$} (outdk.north);

        \end{tikzpicture}%
    }    \caption{The NDR-SHKF architecture, illustrating the flow from the whitened innovation inputs through the N-Deep GRU stack, ending in the policy and auxiliary reconstruction heads.}
    \label{fig:diagram}
\end{figure}

To overcome the static forgetting parameter and numerical instability of the conventional Sage-Husa framework, we propose the N-Deep Recurrent Sage-Husa Kalman Filter (NDR-SHKF). As shown in Fig.~\ref{fig:diagram}, this architecture formulates adaptation factor selection as a sequential decision-making problem. A recurrent neural network learns a vector-valued memory attenuation policy, enabling dimension-wise modulation of process and measurement noise covariances. The pipeline standardizes innovation sequences via a whitening transformation for scale invariance, then passes them through a feed-forward encoder into an N-deep GRU stack designed to disentangle short-term sensor anomalies from long-term dynamic trends. A policy head maps the resulting multi-scale latent representations to the attenuation vector, while an auxiliary decoder reconstructs the input features to regularize the optimization.

\subsection{Whitened Multi-Scale Inputs}

In data-driven filtering, providing neural architectures with absolute measurements ($\mathbf{z}_k$) forces the network to process non-stationary coordinates tied to specific trajectories. This frequently induces overfitting to deterministic spatial paths rather than capturing the underlying stochastic noise profiles~\cite{revach2022kalmannet}. While utilizing predictive residuals mitigates this by isolating the true statistical discrepancies, standard residual representations remain mathematically inadequate. Raw innovation vectors ($\boldsymbol{\nu}_k$) are vulnerable to scaling disparities across heterogeneous sensor modalities, whereas the scalar Normalized Innovation Squared (NIS)~\cite{chen2018weak} collapses the multidimensional error geometry, thereby discarding the directional context essential for identifying specific sensor anomalies.

To provide the neural network with a comprehensive and numerically stable representation of the filter state, a scale-invariant whitening transformation derived from the innovation covariance matrix, $\mathbf{S}_k$, is introduced. At each discrete time step, the lower-triangular Cholesky decomposition of the expected measurement uncertainty is computed:
\begin{equation}
    \mathbf{L}_k \mathbf{L}_k^\top = \mathbf{S}_k + \varepsilon \mathbf{I},
\end{equation}
where $\varepsilon$ denotes a small positive constant added to ensure numerical stability. The innovation vector is subsequently whitened by solving the corresponding triangular system:
\begin{equation}
    \tilde{\boldsymbol{\nu}}_k = \mathbf{L}_k^{-1} \boldsymbol{\nu}_k.
\end{equation}
This operation maps the residual into a standardized Mahalanobis space, quantifying the measurement deviation relative to the predictive standard deviation. Consequently, the network architecture becomes agnostic to the absolute physical units of the underlying sensors.

To preserve the network's awareness of the network regarding the absolute scale of the filter uncertainty, the logarithmic diagonal of the Cholesky factor is extracted:
\begin{equation}
    \mathbf{l}_k = \log(\text{diag}(\mathbf{L}_k) + \varepsilon).
\end{equation}

Furthermore, the full Kalman gain matrix, $\mathbf{K}_k$, is incorporated into the input feature vector via vectorization. The final input feature vector supplied to the network, $\mathbf{y}_k \in \mathbb{R}^{2n_z + n_x n_z}$, is constructed through the concatenation of the whitened innovation, the log-scale uncertainty, and the vectorized Kalman gain:
\begin{equation}
    \mathbf{y}_k = \begin{bmatrix} \tilde{\boldsymbol{\nu}}_k^\top, & \mathbf{l}_k^\top, & \text{vec}(\mathbf{K}_k)^\top \end{bmatrix}^\top.
\end{equation}
To ensure stable gradient propagation during backpropagation and to prevent outlier-induced saturation within the neural layers, $\mathbf{y}_k$ is strictly bounded via numerical clipping.

\subsection{N-Deep Recurrent Context Encoding}\label{subsec:ndrce}

Dynamic tracking scenarios operate across multiple temporal scales: impulsive sensor noise manifests as high-frequency, uncorrelated noise, while continuous maneuvers present as low-frequency, correlated trends. To disentangle these phenomena, the clipped feature vector $\mathbf{y}_k$ is first processed by a feed-forward encoder, parameterized as a Multi-Layer Perceptron (MLP) denoted by $\text{enc}(\cdot)$, to extract nonlinear correlations:

\begin{equation}
    \mathbf{e}_k = \text{enc}(\mathbf{y}_k).
\end{equation}

The encoded feature $\mathbf{e}_k \in \mathbb{R}^{d_e}$ is then fed into a hierarchical stack of $N$ Gated Recurrent Unit (GRU) layers. The hidden states $\mathbf{h}_k^{(i)} \in \mathbb{R}^{d_h}$ for layers $i = 1, \dots, N$ are updated recursively:
\begin{align}
    \mathbf{h}_k^{(1)} &= \text{GRU}^{(1)}(\mathbf{e}_k, \mathbf{h}_{k-1}^{(1)}), \\
    \mathbf{h}_k^{(i)} &= \text{GRU}^{(i)}(\mathbf{h}_k^{(i-1)}, \mathbf{h}_{k-1}^{(i)}) \quad \text{for } i \in[2, N].
\end{align}

To grant the policy simultaneous access to short-term reactive features and long-term historical abstractions, the architecture employs a bifurcated feature routing strategy. The hidden state of the shallowest recurrent layer, $\mathbf{h}_k^{(1)}$, is projected via a fully connected layer, denoted as $\text{ctx}(\cdot)$, into a latent context vector $\mathbf{c}_k = \text{ctx}(\mathbf{h}_k^{(1)})$ to capture immediate, high-frequency anomalies. Conversely, the deepest hidden state, $\mathbf{h}_k^{(N)}$, encodes the long-horizon maneuvering trends. These two scales are subsequently concatenated to form the final policy embedding:
\begin{equation}
    \mathbf{p}_k = \begin{bmatrix} \mathbf{c}_k^\top, & \mathbf{h}_k^{(N)\top} \end{bmatrix}^\top.
\end{equation}

\subsection{Policy Head and Auxiliary Decoder}\label{subsec:pandd}
The multi-scale latent embedding $\mathbf{p}_k$ is routed to the primary \textit{Policy Head} to synthesize the dynamic adaptation factors. The transformation is governed by $\pi(\cdot)$, which represents the core policy network implemented as a feed-forward MLP. A sigmoid activation function, $\sigma(\cdot)$, constrains the outputs to the open interval $(0, 1)$, ensuring that each adapted covariance element remains a convex combination of its prior and empirical estimates. In contrast to classical formulations that rely on a singular parameter, the proposed network generates a multi-dimensional adaptation vector $\mathbf{d}_k \in \mathbb{R}^{n_x + n_z}$:
\begin{equation}
    \mathbf{d}_k = \sigma(\pi(\mathbf{p}_k)).
\end{equation}
In the NDR-SHKF, the adapted process and measurement noise covariances are constrained to diagonal form, with $\mathbf{R}_k = \mathbf{r}_k \mathbf{I}$ and $\mathbf{Q}_k = \mathbf{q}_k \mathbf{I}$ fully parameterized by their diagonal vectors $\mathbf{r}_k \in \mathbb{R}^{n_z}$ and $\mathbf{q}_k \in \mathbb{R}^{n_x}$. Similarly, the empirical estimates $\hat{\mathbf{R}}_{k}$ and $\hat{\mathbf{Q}}_{k}$ defined in Section~\ref{subsec:shkf} are reduced to their diagonals $\hat{\mathbf{r}}_k \in \mathbb{R}^{n_z}$ and $\hat{\mathbf{q}}_k \in \mathbb{R}^{n_x}$. The adapted diagonals are then computed as:
\begin{align}
    \mathbf{r}_{k} &= (\mathbf{1} - \mathbf{d}^{R}_{k}) \odot \mathbf{r}_{k-1} + \mathbf{d}^{R}_{k} \odot \hat{\mathbf{r}}_{k},\\
    \mathbf{q}_{k} &= (\mathbf{1} - \mathbf{d}^{Q}_{k}) \odot \mathbf{q}_{k-1} + \mathbf{d}^{Q}_{k} \odot \hat{\mathbf{q}}_{k},
\end{align}
where $\odot$ denotes the Hadamard product. This diagonal constraint mitigates the risk of non-positive-definite matrices, a frequent failure mode in SHKF caused by finite-sample fluctuations and linearization errors. Diagonalization, symmetrization and positive floor clamping are applied at every recursive step, the same safeguards as in the classical SHKF.

An \textit{Auxiliary Decoder} is employed during the training phase to reconstruct the original whitened input features. This decoding mechanism, parameterized as a multi-layer feed-forward network $\text{dec}(\cdot)$, functions as an inverse mapping that projects the compressed latent representation back into the high-dimensional input space. It operates exclusively on the shallow context vector $\mathbf{c}_k$:
\begin{equation}
    \hat{\mathbf{y}}_k = \text{dec}(\mathbf{c}_k).
\end{equation}
This architectural design fulfills a regularization objective. Heavily parameterized recursive estimators are inherently susceptible to feature collapse, a phenomenon in which the neural policy converges to a static, conservative output for $\mathbf{d}_k$, thereby disregarding the incoming innovation sequence. By requiring the shallow context $\mathbf{c}_k$ to actively reconstruct $\mathbf{y}_k$, the framework effectively enforces that the latent representation preserves the fundamental geometric and statistical properties of the immediate measurement space. This structural decoupling isolates the deeper recurrent layers ($N > 1$), permitting them to specialize exclusively in long-horizon temporal policy synthesis without being constrained by the necessity of high-fidelity input reconstruction.

\subsection{End-to-End Differentiable Optimization}\label{subsec:end2end}

We formulate the entire NDR-SHKF, comprising the EKF prediction, neural feature extraction, policy synthesis, and Sage-Husa covariance updates, as a unified differentiable computational graph. Let $\mathcal{F}_\theta$ denote the single-step NDR-SHKF update function, parameterized by $\theta$. We can view the complete filter state at time step $k$ as the tuple $\mathbf{s}_k = (\hat{\mathbf{x}}_{k|k}, \mathbf{P}_{k|k}, \mathbf{Q}_k, \mathbf{R}_k)$. The recursive evolution of the filter is then expressed as:
\begin{equation}
    \mathbf{s}_k = \mathcal{F}_\theta(\mathbf{s}_{k-1}, \mathbf{z}_k).
\end{equation}

To optimize the meta-policy, we compute the gradient of the cumulative loss $\mathcal{L}_{\text{total}}$ with respect to the network parameters. Because the parameters $\theta$ influence the entire trajectory, optimization is performed via backpropagation Through Time (BPTT)~\cite{werbos1990backpropagation}. Applying the chain rule, the total analytical gradient is the accumulation of gradients at each time step:
\begin{equation}
    \frac{d\mathcal{L}_{\text{total}}}{d\theta} = \sum_{k=1}^T \frac{\partial \mathcal{L}_k}{\partial \mathbf{s}_k} \frac{d\mathbf{s}_k}{d\theta},
\end{equation}
where the total state derivative $\frac{d\mathbf{s}_k}{d\theta}$ is computed recursively:
\begin{equation}
    \frac{d\mathbf{s}_k}{d\theta} = \frac{\partial \mathcal{F}_\theta}{\partial \theta} + \frac{\partial \mathcal{F}_\theta}{\partial \mathbf{s}_{k-1}} \frac{d\mathbf{s}_{k-1}}{d\theta}.
\end{equation}

The optimization objective is formulated as a multi-task loss function evaluated over a trajectory of length $T$. The primary task minimizes the \textit{a posteriori} state estimation error $\hat{\mathbf{x}}_{k|k}$, relative to the ground truth, $\mathbf{x}_k^{\text{true}}$, while the secondary task enforces the auxiliary reconstruction penalty:
\begin{equation}
    \mathcal{L}_{\text{total}}(\theta) = \sum_{k=1}^T \left\| \mathbf{x}_k^{\text{true}} - \hat{\mathbf{x}}_{k|k}(\theta) \right\|_2^2 + \lambda_{\text{aux}} \left\| \mathbf{y}_k - \hat{\mathbf{y}}_k(\theta) \right\|_2^2,
    \label{eq:loss_function}
\end{equation}
where $\lambda_{\text{aux}}$ is a scalar weight balancing the reconstruction penalty.

The constituent operations of the NDR-SHKF, including the Cholesky decomposition of the innovation covariance~\cite{murray2016differentiation}, triangular matrix solves, and matrix inversions for the Kalman gain, are differentiable almost everywhere. By unrolling this computational graph, the gradients can flow from the final tracking error, backward through the recursive SHKF update logic, and into the weights of the N-Deep RNN. 

\begin{algorithm}[t!]
\caption{NDR-SHKF: Single-Step Recursive Update}
\label{alg:ndrshkf}
\begin{algorithmic}
\REQUIRE $\hat{\mathbf{x}}_{k-1|k-1}$, $\mathbf{P}_{k-1|k-1}$, $\mathbf{q}_{k-1}$, $\mathbf{r}_{k-1}$, $\mathbf{z}_k$, $\{\mathbf{h}^{(i)}_{k-1}\}_{i=1}^{N}$
\vspace{0.25cm}
\STATE \textbf{Prediction \& Kalman Gain}
\STATE $\mathbf{F}_{k-1} \gets \left.\frac{\partial f}{\partial \mathbf{x}}\right|_{\hat{\mathbf{x}}_{k-1|k-1}}$
\STATE $\hat{\mathbf{x}}_{k|k-1} \gets f(\hat{\mathbf{x}}_{k-1|k-1})$, \quad $\mathbf{P}_{k|k-1} \gets \mathbf{F}_{k-1}\mathbf{P}_{k-1|k-1}\mathbf{F}_{k-1}^\top + \mathbf{q}_{k-1}\mathbf{I}$
\STATE $\mathbf{H}_k \gets \left.\frac{\partial h}{\partial \mathbf{x}}\right|_{\hat{\mathbf{x}}_{k|k-1}}$
\STATE $\boldsymbol{\nu}_k \gets \mathbf{z}_k - h(\hat{\mathbf{x}}_{k|k-1})$, \quad $\mathbf{S}_k \gets \mathbf{H}_k\mathbf{P}_{k|k-1}\mathbf{H}_k^\top + \mathbf{r}_{k-1}\mathbf{I}$
\STATE $\mathbf{K}_k \gets \mathbf{P}_{k|k-1}\mathbf{H}_k^\top\mathbf{S}_k^{-1}$
\vspace{0.25cm}
\STATE \textbf{Feature Construction}
\STATE $\mathbf{L}_k \gets \operatorname{Cholesky}(\mathbf{S}_k + \varepsilon\mathbf{I})$, \quad $\tilde{\boldsymbol{\nu}}_k \gets \mathbf{L}_k^{-1}\boldsymbol{\nu}_k$
\STATE $\mathbf{y}_k \gets \operatorname{clip}\!\left(\begin{bmatrix}\tilde{\boldsymbol{\nu}}_k^\top,\ \log\!\left(\operatorname{diag}(\mathbf{L}_k) + \varepsilon\right)^\top,\ \operatorname{vec}(\mathbf{K}_k)^\top\end{bmatrix}^\top\right)$

\vspace{0.35cm}
\STATE \textbf{N-Deep GRU Encoding \& Policy Synthesis}
\STATE $\mathbf{e}_k \gets \operatorname{enc}(\mathbf{y}_k)$
\STATE $\mathbf{h}_k^{(1)} \gets \operatorname{GRU}^{(1)}(\mathbf{e}_k,\ \mathbf{h}_{k-1}^{(1)})$
\FOR{$i = 2$ \textbf{to} $N$}
    \STATE $\mathbf{h}_k^{(i)} \gets \operatorname{GRU}^{(i)}(\mathbf{h}_k^{(i-1)},\ \mathbf{h}_{k-1}^{(i)})$
\ENDFOR
\STATE $\mathbf{c}_k \gets \operatorname{ctx}(\mathbf{h}_k^{(1)})$, 
\STATE $\mathbf{p}_k \gets \begin{bmatrix}\mathbf{c}_k^\top,\ \mathbf{h}_k^{(N)\top}\end{bmatrix}^\top$
\STATE  $[\mathbf{d}_k^{Q},\ \mathbf{d}_k^{R}] \gets \operatorname{partition}\!\left(\sigma\!\left(\pi(\mathbf{p}_k)\right)\right)$ 
\vspace{0.05cm}
\STATE $\hat{\mathbf{y}}_k \gets \operatorname{dec}(\mathbf{c}_k)$ \hfill $\triangleright$ \textit{Reconstruction omitted during inference.}

\vspace{0.35cm}
\STATE \textbf{State \& Adaptive Covariance Updates}
\STATE $\hat{\mathbf{x}}_{k|k} \gets \hat{\mathbf{x}}_{k|k-1} + \mathbf{K}_k\boldsymbol{\nu}_k$
\STATE $\mathbf{P}_{k|k} \gets (\mathbf{I} - \mathbf{K}_k\mathbf{H}_k)\mathbf{P}_{k|k-1}$
\vspace{0.1cm}
\STATE $\hat{\mathbf{r}}_k \gets \text{diag}(\boldsymbol{\nu}_k\boldsymbol{\nu}_k^\top - \mathbf{H}_k\mathbf{P}_{k|k-1}\mathbf{H}_k^\top)$
\vspace{0.1cm}
\STATE $\hat{\mathbf{q}}_k \gets \text{diag}( \mathbf{K}_k\boldsymbol{\nu}_k\boldsymbol{\nu}_k^\top\mathbf{K}_k^\top + \mathbf{P}_{k|k} - \mathbf{F}_{k-1}\mathbf{P}_{k-1|k-1}\mathbf{F}_{k-1}^\top)$
\vspace{0.1cm}
\STATE $\mathbf{r}_{k} = (\mathbf{1} - \mathbf{d}^{R}_{k}) \odot \mathbf{r}_{k-1} + \mathbf{d}^{R}_{k} \odot \hat{\mathbf{r}}_{k}$
\vspace{0.1cm}
\STATE $\mathbf{q}_{k} = (\mathbf{1} - \mathbf{d}^{Q}_{k}) \odot \mathbf{q}_{k-1} + \mathbf{d}^{Q}_{k} \odot \hat{\mathbf{q}}_{k}$
\RETURN $\hat{\mathbf{x}}_{k|k}$, $\mathbf{P}_{k|k}$, $\mathbf{q}_k$, $\mathbf{r}_k$, $\{\mathbf{h}_k^{(i)}\}_{i=1}^{N}$
\end{algorithmic}
\end{algorithm}

\subsection{Computational Complexity Analysis}\label{subsec:complexity}
We analyze the asymptotic computational complexity of the NDR-SHKF during inference to evaluate its viability for real-time deployment on embedded UAV hardware. Let $n_x$ denote the state dimension, $n_z$ the measurement dimension, $N$ the recurrent network depth, and $d_h$ the hidden dimension. The execution time per step comprises three phases: classical recursive estimation, input whitening, and the neural hyper-controller forward pass.

First, EKF operations establish the baseline computational load. State covariance prediction requires $\mathcal{O}(n_x^3)$ operations. Kalman gain computation involves matrix inversion ($\mathcal{O}(n_z^3)$) and multiplication ($\mathcal{O}(n_x^2 n_z + n_x n_z^2)$). Traditional Sage-Husa empirical updates add an $\mathcal{O}(n_x^2 + n_z^2)$ overhead \cite{cheng2024online}. The classical algorithmic complexity is therefore bounded by $\mathcal{O}(n_x^3 + n_z^3 + n_x^2 n_z + n_x n_z^2)$.

Second, standardizing the innovation vector requires a Cholesky decomposition of $\mathbf{S}_k$ ($\mathcal{O}(n_z^3)$) and a lower-triangular solve ($\mathcal{O}(n_z^2)$). Because $\mathbf{S}_k$ is assembled during the standard update phase, this introduces minimal overhead and remains within the existing $\mathcal{O}(n_z^3)$ ceiling.

Finally, the neural hyper-controller processes an input vector of dimension $2n_z + n_x n_z$. The initial encoder requires $\mathcal{O}((2 n_z + n_x n_z) d_{enc} + d_{enc} d_e)$ operations, where $d_{enc}$ denotes the encoder first hidden dimension. The $N$-layer GRU stack scales as $\mathcal{O}(d_e d_h + N d_h^2)$, the policy head, $\pi(\cdot)$, a two-layer MLP with hidden dimension $d_\pi$, requires $\mathcal{O}(d_\pi(d_p + d_h + n_x + n_z))$ operations for $N > 1$, or $\mathcal{O}(d_\pi(d_p + n_x + n_z))$ for $N = 1$. The auxiliary decoder is omitted during inference. 

Aggregating these phases and dropping dominated terms yields the total asymptotic time complexity per discrete time step:
\begin{equation}
    \mathcal{O}\Big(n_x^3 + n_z^3 + n_x^2 n_z + n_x n_z^2 + n_x n_z d_{enc} + N d_h^2\Big).
\end{equation}

\section{Out-of-Distribution Generalization Across Chaotic Dynamical Systems}\label{sec:sim_experiments}
A central limitation of purely data-driven estimators is their tendency to overfit the training domain, often causing divergence when the underlying process dynamics or measurement mappings are perturbed. To evaluate the Out-of-Distribution (OOD) generalization of the NDR-SHKF, we present an OOD transfer experiment across topologically distinct chaotic attractors.

\subsection{Training Domain: Lorenz Attractor}\label{subsec:lorenz_attractor}
During the training phase, the recurrent policy was optimized exclusively utilizing simulated trajectories from a three-dimensional Lorenz chaotic attractor~\cite{lorenz1963deterministic}. The continuous-time process dynamics are governed by the coupled differential equations:
\begin{align}
    \dot{x} &= \sigma_{L}(y - x), \\
    \dot{y} &= x(\rho_{L} - z) - y, \\
    \dot{z} &= xy - \beta_{L} z,
\end{align}
parameterized by the standard coefficients $\sigma_{L}=10.0$, $\rho_{L}=28.0$, and $\beta_{L}=8/3$~\cite{strogatz2015nonlinear}. These dynamics are numerically integrated utilizing a fourth-order Runge-Kutta scheme. To ensure coverage of the chaotic manifold, initial states for each episode were independently drawn from uniform distributions: $x_0, y_0 \sim \mathcal{U}(-15, 15)$ and $z_0 \sim \mathcal{U}(10, 40)$. The observation model for this training domain was linear, mapping the state space to extract the $x$ and $z$ positional coordinates: 
\begin{equation}
    \mathbf{z}_k = \begin{bmatrix} x,\  z \end{bmatrix}^\top + \mathbf{v}_k.
\end{equation}

The underlying true process noise covariance was modeled as a time-varying diagonal matrix, where each state dimension is independently modulated with a sinusoidal pulse to emulate unmodeled, axis-specific harmonic vibrations:
\begin{equation}
    \mathbf{Q}_k = \text{diag}\big(q_{1}(k),\ q_{2}(k),\ q_{3}(k) \big),
\end{equation}
where each diagonal element is defined as 

\begin{equation}
    q_{i}(k) = q_{\text{base}} \big(1 + A_i \sin^2(\omega_i k \Delta t + \phi_i)\big).
\end{equation}

Simultaneously, the measurement noise $\mathbf{v}_k$ was drawn from a Gaussian Mixture Model (GMM) to simulate measurement outliers:
\begin{equation}
    \mathbf{v}_k \sim (1-\epsilon)\mathcal{N}(\mathbf{0}, \mathbf{R}_{\text{base}}) + \epsilon\mathcal{N}(\mathbf{0}, \eta \mathbf{R}_{\text{base}}),
\end{equation}
where $\epsilon$ represents the probability of an outlier, and $\eta$ is the outlier scaling factor. During the Lorenz training phase, the environment exhibited axis-specific process perturbations with a baseline noise scalar of $q_{\text{base}} = 0.01$. To ensure the network generalizes across diverse dynamic conditions, the harmonic parameters for each dimension were independently sampled from uniform distributions at the beginning of each episode: amplitudes $A_i \sim \mathcal{U}(0, 0.2)$, frequencies $\omega_i \sim \mathcal{U}(0.1, 1.0)$ rad/s, and phase shifts $\phi_i \sim \mathcal{U}(0, 2\pi)$. To break dimensional symmetry in the observation space, the linear sensors were subjected to a non-uniform nominal covariance of $\mathbf{R}_{\text{base}} = \text{diag}(1.0, 2.0)$, an outlier probability of $\epsilon = 0.05$ (5\%), and an outlier scale factor of $\eta = 5$.

To accommodate the 3-dimensional state space ($n_x=3$) and 2\text{-}dimensional measurement space ($n_z=2$), the NDR-SHKF feature vector was dimensioned to $\mathbf{y}_k \in \mathbb{R}^{10}$. The feed-forward encoder, $\text{enc}(\cdot)$, comprises a two-layer MLP with $d_{enc}$ and $d_e$ neurons (configured as 32 and 16, respectively), utilizing rectified linear unit (ReLU) activations~\cite{glorot2011deep}. The recurrent temporal core was evaluated at varying architectural depths ($N \in \{1, 3, 5\}$), maintaining a constant hidden state capacity of $d_h=32$ across all GRU cells. The shallow context projection, $\text{ctx}(\cdot)$, utilizes a two-layer MLP to map the primary hidden state through a 32-neuron hidden layer into a 32-dimensional latent representation and a final ReLU activation. From this context, the auxiliary decoder, $\text{dec}(\cdot)$, reconstructs the 10-dimensional input via a two-layer network featuring hidden layers of 16 and 32 neurons. Finally, the policy head, $\pi(\cdot)$, is configured as a two-layer MLP with two 16-neuron hidden layers. It maps the multi-scale embedding (dimension 64 for $N>1$, or 32 for $N=1$), followed by the bounding sigmoid activation, down to the 5-dimensional adaptation vector $\mathbf{d}_k \in \mathbb{R}^{5}$.

The NDR-SHKF models were trained end-to-end via BPTT described in Section~\ref{subsec:end2end}. Trajectories were simulated in sequences of $T = 60$ discrete time steps ($\Delta t = 0.01$s) with a batch size of 64. Optimization was performed over 1000 epochs utilizing the Adam optimizer~\cite{kingma2014adam} with a learning rate of $10^{-3}$. To maintain gradient stability, global gradient norm clipping was enforced at a threshold of $0.5$. The multi-task loss formulation utilized a reconstruction weight of $\lambda_{\text{aux}} = 0.1$.

\subsection{Out-of-Distribution Testing: R{\"o}ssler Attractor}
\begin{figure}[t!]
    \centering
    \includegraphics[width=1.0\columnwidth]{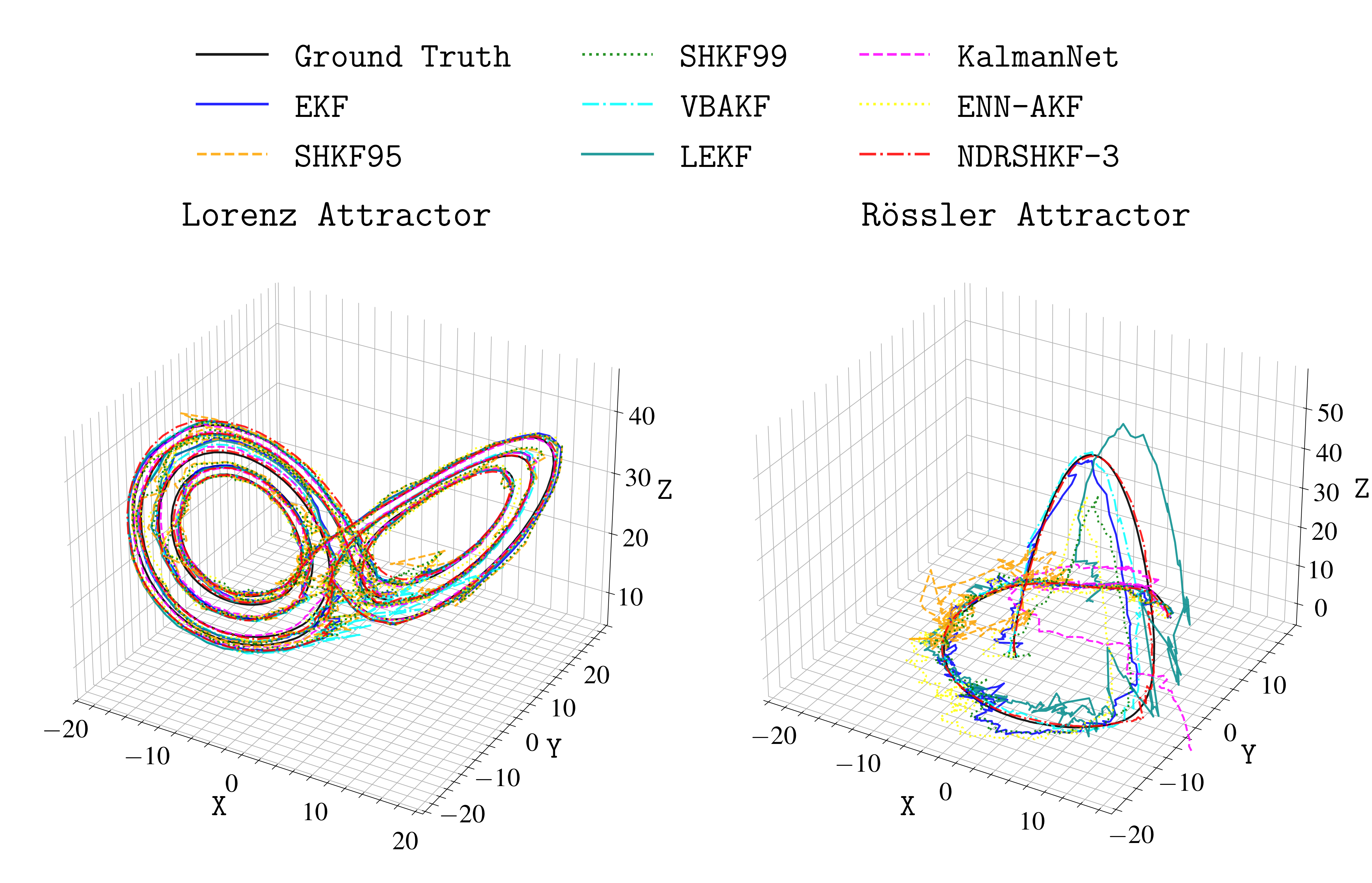} 
    \caption{Side-by-side 3D state trajectories of the Lorenz and R{\"o}ssler attractors. The plots compare the ground truth manifold against the estimates produced by the EKF, various adaptive and learning-based baselines, and the proposed NDR-SHKF architectures under heavy-tailed noise conditions.}
    \label{fig:attractors}
\end{figure}

Following convergence, the network weights were fixed, and the policy was evaluated against the R{\"o}ssler attractor~\cite{rossler1976equation}. As illustrated in Fig.~\ref{fig:attractors}, this system presents a distinct spatial topology, characterized by the following set of nonlinear differential equations: 
\begin{align}
    \dot{x} &= -y - z, \\
    \dot{y} &= x + a_{R}y, \\
    \dot{z} &= b_{R} + z(x - c_{R}),
\end{align}
parameterized by $a_{R}=0.2$, $b_{R}=0.2$, and $c_{R}=5.7$. Initial system states were sampled from: $x_0, y_0 \sim \mathcal{U}(-10, 10)$ and $z_0 \sim \mathcal{U}(0, 10)$. To further test the OOD condition, the linear observation model was replaced with a nonlinear, radar-inspired range-and-bearing formulation outputting Euclidean radial distance and azimuth angle:
\begin{equation}
    \mathbf{z}_k = \begin{bmatrix} \sqrt{x^2+y^2},\  \arctan(y, x) \end{bmatrix}^\top + \mathbf{v}_k.
\end{equation}

To evaluate the model's capacity for cross-domain generalization, the stochastic parameters governing the noise distributions were purposefully shifted from the training baseline. In this testing domain, the process noise oscillation amplitude distribution was increased to $A_i \sim \mathcal{U}(0, 1.0)$, the outlier probability was increased to $\epsilon = 0.10$ (10\%), accompanied by an outlier scale factor of $\eta = 10$.

\subsection{Benchmarked Baselines}\label{subsec:baselines}

To evaluate the cross-domain generalization capabilities of the proposed NDR-SHKF, it is compared against classical, adaptive, and learning-based filtering algorithms. For a consistent methodology, all estimators were identically initialized with the nominal covariance matrices $\mathbf{Q}_{\text{base}} = q_{\text{base}} \mathbf{I}_{3}$ and $\mathbf{R}_{\text{base}} = \text{diag}(1.0, 2.0)$, requiring them to detect and adapt to unmodeled variances through the real-time innovation sequence. The evaluated baselines are configured as follows:

\begin{itemize}
    \item \textbf{EKF:} The standard non-adaptive Extended Kalman Filter, operating with the static, initialized $\mathbf{Q}$ and $\mathbf{R}$ matrices.

    \item \textbf{SHKF95 \& SHKF99:} The conventional Sage-Husa Kalman Filter employing static scalar forgetting factors of $b=0.95$ and $b=0.99$, respectively. These represent the classical trade-off between transient responsiveness and steady-state noise attenuation.

    \item \textbf{VBAKF \cite{huang2017novel}:} The Variational Bayesian Adaptive Kalman Filter, which approximates unknown measurement noise distributions using an Inverse-Wishart prior. The algorithm is configured with an initial degrees-of-freedom parameter $\tau=3.0$, a fading factor $\rho = 1 - e^{-4}$, and performs $N_{\text{iter}}=10$ variational iterations per measurement update.

    \item \textbf{ENN-AKF \cite{gao2012seam}:} A hybrid Elman Neural Network Adaptive Kalman Filter. Operating as an external residual compensation module, it employs a recurrent neural network with a 64-dimensional hidden state to ingest past states and innovations, additively correcting the final state estimate of a baseline AKF.

    \item \textbf{L-EKF \cite{zheng2019rnn}:} The Learnable Extended Kalman Filter, which utilizes a recurrent Gain Modification Network (GMN). Driven by a 64-dimensional GRU, the neural policy processes the state estimate and innovation sequence to multiplicatively perturb the analytically derived Kalman gain.    

    \item \textbf{KalmanNet \cite{revach2022kalmannet}:} A data-driven Kalman estimator that circumvents the Riccati covariance recursions by utilizing a 64-dimensional RNN to directly synthesize the Kalman gain matrix from instantaneous measurement innovations and temporal state differences. This implementation employs Architecture 1 from the original formulation. Unlike covariance-based filters, KalmanNet produces a raw, unbounded gain, which can cause state divergence when trajectories deviate from the chaotic Lorenz attractor during early training, resulting in numerical instability. To ensure robustness, explicit numerical safeguards were integrated: state estimates and input features are clipped to $\pm 50$, and the training loss function masks gradient values that exceed defined numerical thresholds.
\end{itemize}

Consistent with the NDR-SHKF, all learning-based baselines (ENN-AKF, L-EKF, and KalmanNet) were trained on the Lorenz attractor utilizing identical domain randomization parameters. Their neural components were optimized using the same BPTT scheme, batch size (64), temporal sequence length ($T=60$), and optimization hyperparameters (Adam optimizer, learning rate $10^{-3}$, gradient clipping at 0.5) to isolate and evaluate the architectural differences of the proposed framework.

\subsection{Evaluation Metrics and OOD Results}
To quantify the tracking performance across the Monte Carlo simulations, we define two primary evaluation metrics. Let $\mathbf{e}_k^{(j)} = \mathbf{x}_k^{(j)} - \hat{\mathbf{x}}_k^{(j)}$ denote the state estimation error vector at discrete time step $k$ for a given Monte Carlo run $j$. The instantaneous spatial Root Mean Square Error (RMSE) across the $n_x$ state dimensions is computed as:
\begin{equation}
    \text{RMSE}_k^{(j)} = \sqrt{ \frac{1}{n_x} \big(\mathbf{e}_k^{(j)}\big)^\top \mathbf{e}_k^{(j)} }.
\end{equation}

The primary scalar metric presented in Table~\ref{tab:results} is the \textbf{Average Root Mean Square Error (ARMSE)}. This metric represents the time-averaged spatial error, subsequently averaged over all $N$ non-divergent Monte Carlo runs to provide a global measure of expected accuracy:
\begin{equation}
    \text{ARMSE} = \frac{1}{N} \sum_{j=1}^{N} \left( \frac{1}{T} \sum_{k=1}^T \text{RMSE}_k^{(j)} \right),
\end{equation}
where $T$ is the total number of simulated time steps.

To visualize the accumulation of error and the long-term stability bounds of the filters, Fig.~\ref{fig:sim_rmse} evaluates the \textbf{Cumulative RMSE (CRMSE)}. Evaluated over the entire non-diverging ensemble up to an arbitrary time step $t$, this metric is formulated as the root-mean-square of all historical spatial errors:
\begin{equation}
    \text{CRMSE}(t) = \sqrt{ \frac{1}{N \cdot t} \sum_{j=1}^{N} \sum_{k=1}^t \big(\text{RMSE}_k^{(j)}\big)^2} 
\end{equation}
Trajectories exhibiting numerical divergence (defined numerically as yielding instantaneous errors exceeding a large outlier threshold of 100 meters, or resulting in numerical instability) were isolated. Their occurrences are reported separately as the "Divergence (\%)" metric to ensure the statistical integrity of the RMSE evaluations.

\subsection{Results and Discussion}

\begin{figure}[t!]
    \centering
    \includegraphics[width=1.0\columnwidth]{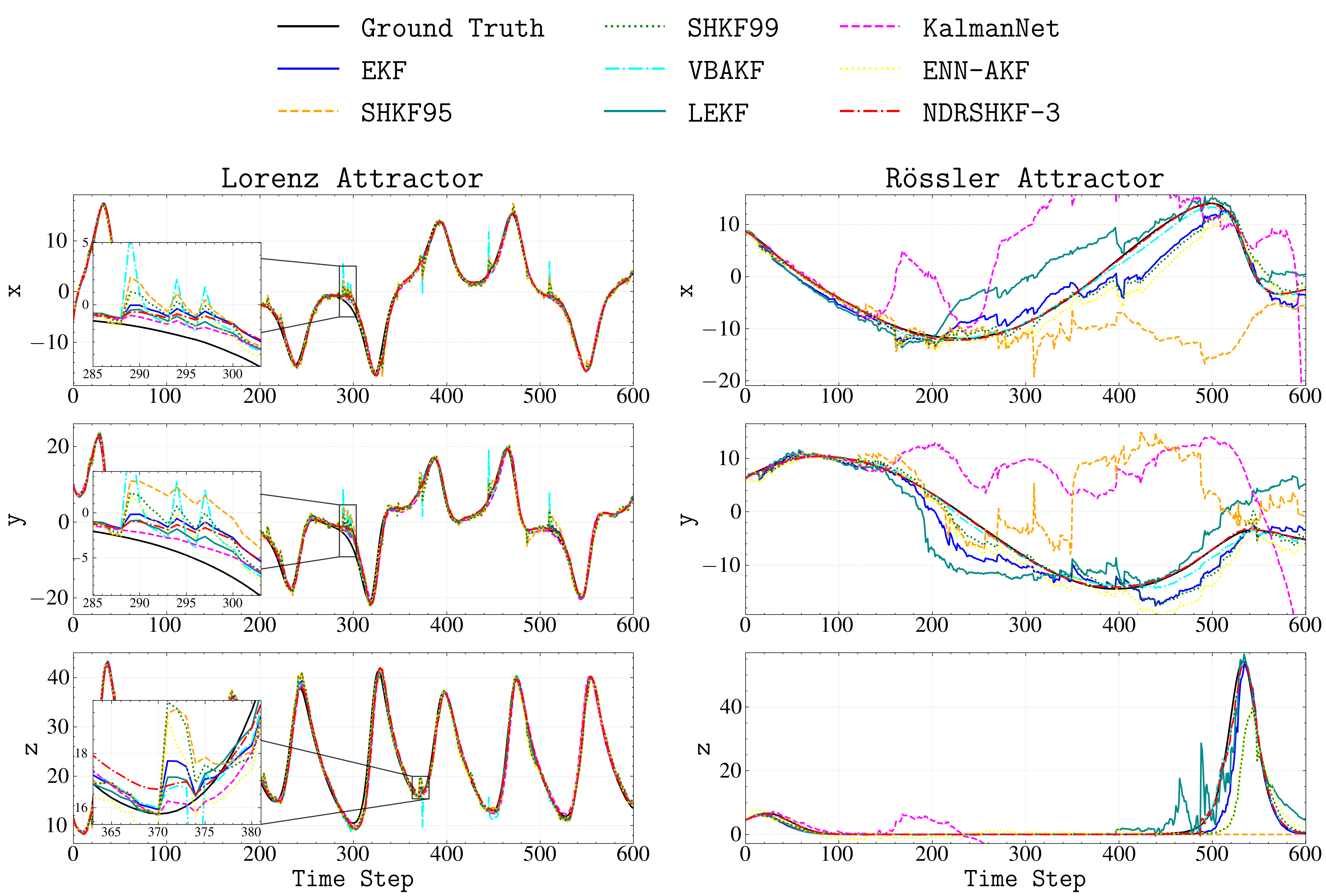} 
    \caption{Coordinate-wise tracking performance ($x, y, z$) over time for both the Lorenz and R{\"o}ssler dynamical systems. The plots compare the ground truth against the proposed NDR-SHKF architectures and various classical, adaptive, and learning-based baselines. The NDR-SHKF rejects instantaneous measurement spikes while maintaining tracking during continuous out-of-distribution maneuvers.}
    \label{fig:xyz_tracking}
\end{figure}

\begin{figure}[t!]
    \centering
    \includegraphics[width=1.0\columnwidth]{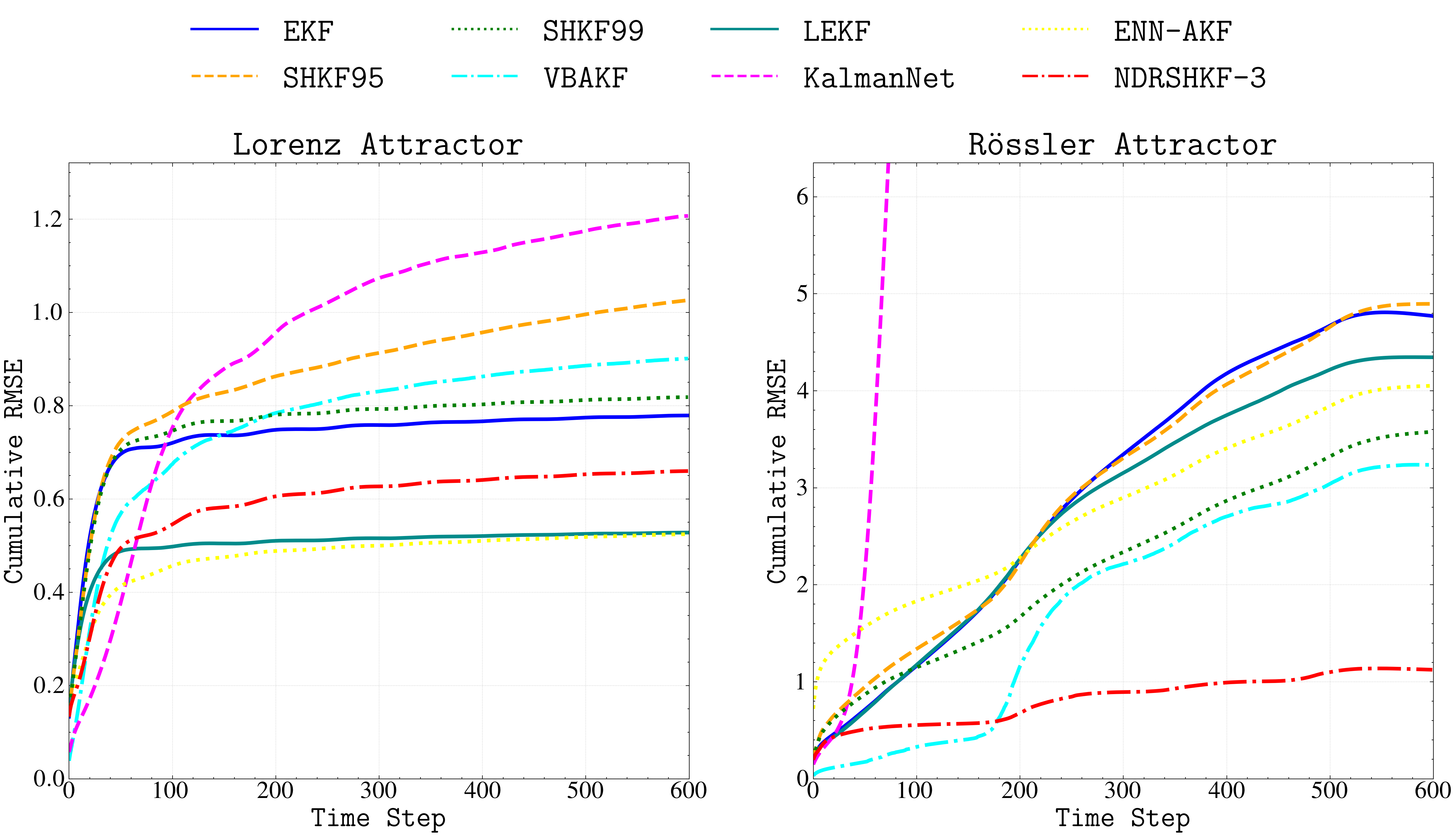} 
    \caption{Cumulative Root Mean Square Error (CRMSE) evaluated over the 600-step test trajectories. The proposed NDR-SHKF architectures demonstrate lower long-term error bounds, particularly in the Out-of-Distribution R{\"o}ssler scenario where data-driven baselines exhibit limited generalization.}
    \label{fig:sim_rmse}
\end{figure}

\begin{table}[t!]
    \centering
    \caption{Monte Carlo Performance Benchmark (10000 Runs, 600 Steps). Divergence represents the percentage of trajectories yielding instantaneous errors exceeding 100 meters or numerical instability. A baseline divergence floor of 2.2\% arises from ground-truth attractor instability on the R{\"o}ssler system and affects all methods equally. Lorenz divergence was 0\% for all methods. The NDR-SHKF entry corresponds to a single independently trained model (seed distinct from the ablation study in Section~\ref{subsec:ablation})}
    \label{tab:results}
    \begin{tabular}{lccc}
        \hline
        \textbf{Method} & \textbf{Lorenz (In-Dist)} & \multicolumn{2}{c}{\textbf{R{\"o}ssler (Out-of-Dist)}} \\
                        & mean $\pm$ std & mean $\pm$ std & Div. (\%) \\ \hline
        EKF & $0.652 \pm 0.108$ & $2.894 \pm 2.146$ & 2.25\% \\
        SHKF95 & $0.838 \pm 0.126$ & $3.118 \pm 1.706$ & 2.71\% \\
        SHKF99 & $0.684 \pm 0.121$ & $2.255 \pm 1.506$ & 2.29\% \\
        VBAKF & $0.641 \pm 0.126$ & $1.172 \pm 1.087$ & 5.37\% \\
        ENN-AKF & $0.428 \pm 0.078$ & $2.845 \pm 1.510$ & 2.29\% \\
        L-EKF & $0.434 \pm 0.066$ & $2.838 \pm 1.768$ & 2.31\% \\
        KalmanNet & $0.694 \pm 0.288$ & $29.792 \pm 10.329$ & 2.23\% \\
        \hline
        NDR-SHKF (3-Layer) & $0.527 \pm 0.126$ & $\mathbf{0.493 \pm 0.548}$ & 2.23\% \\
        \hline
    \end{tabular}
\end{table}%

Table~\ref{tab:results} summarizes the quantitative results of 10,000 Monte Carlo simulations evaluated over 600 time steps. Coordinate-wise tracking performance and Cumulative RMSE are visualized in Fig.~\ref{fig:xyz_tracking} and Fig.~\ref{fig:sim_rmse}, respectively.

Within the nominal Lorenz training domain, the neural-augmented filters successfully captured the underlying noise characteristics, outperforming the classical EKF and SHKF baselines. The data-driven ENN-AKF and L-EKF achieved the lowest absolute tracking errors ($0.428$ and $0.434$, respectively), demonstrating strong adaptation to the training distribution. The NDR-SHKF achieved a competitive in-distribution error of $0.527$, confirming that the constrained adaptation mechanism does not compromise nominal filtering performance.

The cross-domain generalization capability of the NDR-SHKF is most evident in the Out-of-Distribution R{\"o}ssler test. Despite structural shifts in both the temporal evolution of the system and the spatial geometry of the nonlinear measurements, the NDR-SHKF maintained an ARMSE of $0.493$, an order of magnitude lower than the closest classical adaptive baseline (SHKF99 at $2.255$) and substantially outperforming all learning-based alternatives. The ENN-AKF and L-EKF, despite their in-distribution superiority, exhibited overfitting with OOD errors of $2.845$ and $2.838$, respectively. KalmanNet diverged on the R{\"o}ssler system, yielding an ARMSE of $29.792$, confirming that purely data-driven architectures lack the structural stability guarantees inherent to covariance-based filters when confronted with unseen dynamics. The VBAKF achieved moderate R{\"o}ssler performance ($1.172$) but suffered the highest divergence rate ($5.37\%$), nearly double that of any other method. All methods exhibit a baseline divergence rate of approximately $2.2\%$ on the R{\"o}ssler system, of which 2.17 percentage points are attributable to ground-truth trajectory instability rather than filter failure; no method experienced any divergence on the Lorenz attractor.

\subsection{Ablation Study}\label{subsec:ablation}

To isolate the contribution of each architectural component, three ablation axes were evaluated: recurrent depth, auxiliary reconstruction weight, and input representation. All variants were trained under identical conditions to the NDR-SHKF (3-Layer) baseline (Section~\ref{subsec:lorenz_attractor}), modifying only the targeted parameter. Each variant was trained across 100 independent weight initializations and evaluated on a fixed held-out test set comprising 10,000 Monte Carlo trajectories per domain.

\begin{figure}[t!]
    \centering
    \includegraphics[width=1.0\columnwidth]{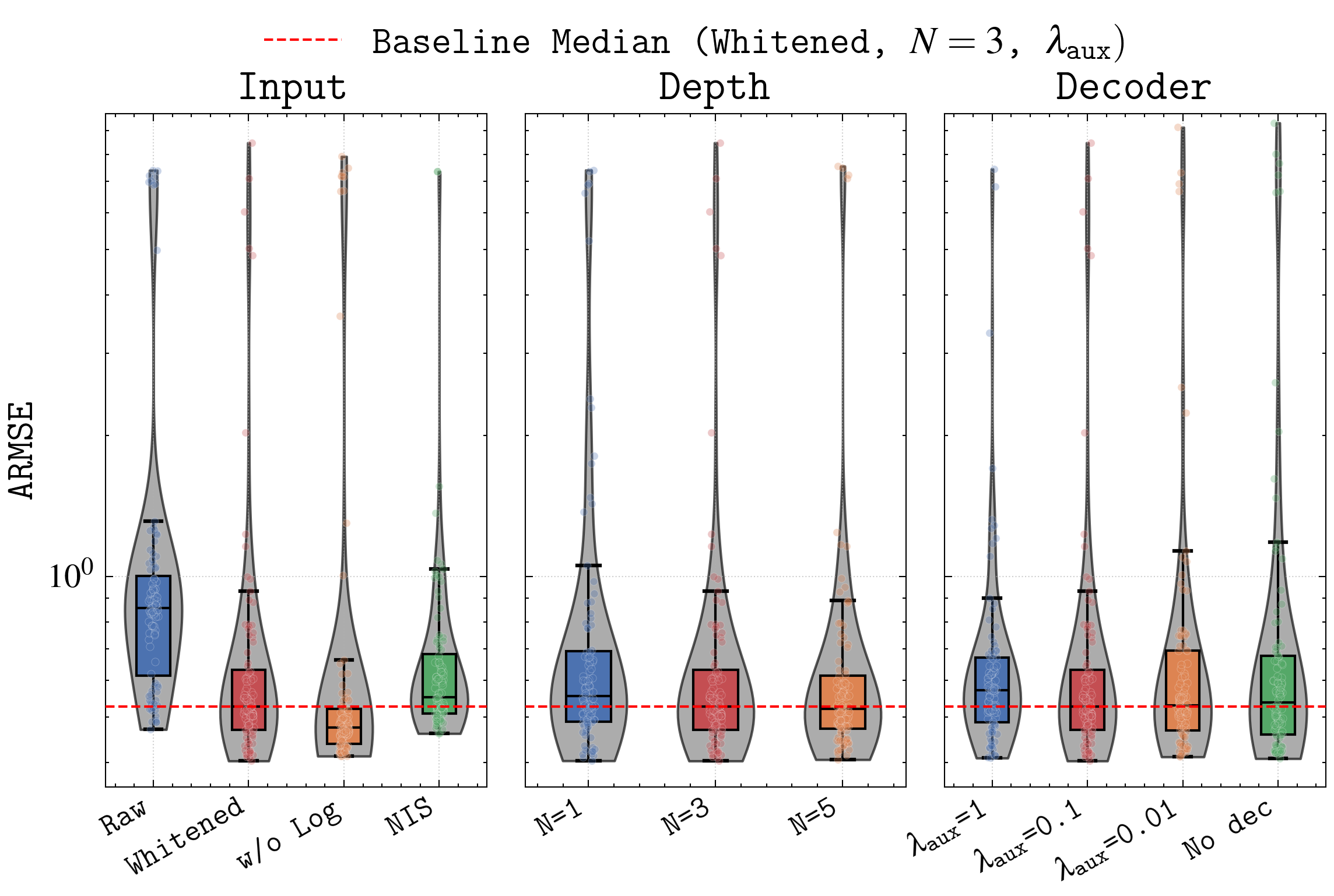} 
    \caption{Distribution of R{\"o}ssler out-of-distribution ARMSE across 100 training seeds for each ablation variant. Violin curves show the kernel density estimate; boxed markers indicate the median; whiskers extend to the 5th and 95th percentiles. In-distribution (Lorenz) distributions are omitted as all variants achieve comparable performance.}
    \label{fig:violin_box}
\end{figure}

Because the ARMSE distributions across random seeds exhibit heavy tails, a small fraction of initializations produce divergent filters (Fig.~\ref{fig:violin_box}). Consequently, we report both the mean (capturing worst-case initialization sensitivity) and the median (representing typical performance). Divergence rates are computed by pooling all trajectories across the 100 seeds ($1{,}000{,}000$ total per domain). 

\subsubsection{Recurrent Depth Ablation}
The depth of the GRU stack governs the temporal receptive field (Section~\ref{subsec:ndrce}). Table~\ref{tab:ablation_depth} reports performance for $N \in \{1, 3, 5\}$.

\begin{table}[t!]
    \centering
    \caption{Effect of recurrent depth $N$ on the NDR-SHKF. Mean $\pm$ std and median ARMSE across 100 training seeds.}
    \label{tab:ablation_depth}
    \small
    \begin{tabular}{lccc}
        \hline
        \textbf{Depth} & \textbf{Lorenz} & \multicolumn{2}{c}{\textbf{R{\"o}ssler (OOD)}} \\
                       & mean $\pm$ std [med] & mean $\pm$ std [med] & Div. \\ \hline
        1-Layer & $0.698 \pm 0.692$ [$0.521$] & $1.028 \pm 1.495$ [$0.555$] & 4.09\% \\
        3-Layer & $0.661 \pm 0.648$ [$0.528$] & $0.872 \pm 1.299$ [$0.527$] & \textbf{3.57\%} \\
        5-Layer & $0.666 \pm 0.691$ [$0.522$] & $\mathbf{0.843 \pm 1.337}$ [$\mathbf{0.521}$] & 3.95\% \\ \hline
    \end{tabular}
\end{table}

The 5-layer architecture achieves the lowest mean ($0.843$) and median ($0.521$), confirming that deeper temporal abstractions enhance OOD generalization. However, the 3-layer architecture achieves competitive accuracy ($0.527$ median) while minimizing the divergence rate ($3.57\%$), offering the favorable trade-off between representational capacity and training stability. The 1-layer variant fails to generalize effectively, exhibiting the highest median error ($0.555$) and divergence ($4.09\%$).

\subsubsection{Auxiliary Decoder Ablation}
The auxiliary reconstruction objective prevents policy degeneration by enforcing the shallow context $\mathbf{c}_k$ to retain sufficient information (Section~\ref{subsec:pandd}). Table~\ref{tab:ablation_decoder} evaluates regularization strengths $\lambda_{\text{aux}} \in \{1.0, 0.1, 0.01, 0.0\}$.

\begin{table}[t!]
    \centering
    \caption{Effect of the auxiliary reconstruction objective ($\lambda_{\text{aux}}$). Mean $\pm$ std and median ARMSE across 100 training seeds.}
    \label{tab:ablation_decoder}
    \small
    \begin{tabular}{lccc}
        \hline
        \textbf{Method} & \textbf{Lorenz} & \multicolumn{2}{c}{\textbf{R{\"o}ssler (OOD)}} \\
                        & mean $\pm$ std [med] & mean $\pm$ std [med] & Div. \\ \hline
        $\lambda_{\text{aux}} = 1.0$ & $0.600 \pm 0.456$ [$0.531$] & $\mathbf{0.774 \pm 0.972}$ [$0.570$] & $\mathbf{3.08\%}$ \\
        $\lambda_{\text{aux}} = 0.1$ (Proposed) & $0.661 \pm 0.648$ [$0.528$] & $0.872 \pm 1.299$ [$\mathbf{0.527}$] & 3.57\% \\
        $\lambda_{\text{aux}} = 0.01$ & $0.663 \pm 0.708$ [$0.530$] & $0.887 \pm 1.397$ [$0.530$] & 3.91\% \\
        $\lambda_{\text{aux}} = 0.0$ (No Decoder) & $0.719 \pm 0.753$ [$0.534$] & $1.045 \pm 1.700$ [$0.538$] & 4.81\% \\ \hline
    \end{tabular}
\end{table}

Removing the decoder ($\lambda_{\text{aux}} = 0.0$) yields the highest divergence ($4.81\%$) and worst mean ARMSE ($1.045$), confirming the architecture is susceptible to feature collapse without explicit regularization. Conversely, maximum regularization ($\lambda_{\text{aux}} = 1.0$) maximizes stability (divergence $3.08\%$) but degrades typical performance (median increases to $0.570$). The proposed $\lambda_{\text{aux}} = 0.1$ balances these constraints, achieving the lowest median OOD error while reducing divergence relative to the unregularized baseline.

\subsubsection{Input Representation Ablation}
Table~\ref{tab:ablation_input_form} formulates four alternative input representations to isolate the contributions of the whitening transformation and the logarithmic uncertainty magnitude. Performance is reported in Table~\ref{tab:ablation_input_perf}.

\begin{table}[t!]
    \centering
    \caption{Input representation variants. Each modifies $\mathbf{y}_k$ while retaining the 3-layer architecture.}
    \label{tab:ablation_input_form}
    \renewcommand{\arraystretch}{1.4}
    \small
    \begin{tabular}{lcc}
        \hline
        \textbf{Variant} & \textbf{Input $\mathbf{y}_k$} & \textbf{Dim.} \\ \hline
        Whitened (Proposed) & $\begin{bmatrix} \mathbf{L}_k^{-1}\boldsymbol{\nu}_k^\top, & \log(\text{diag}(\mathbf{L}_k)\!+\!\varepsilon)^\top, & \text{vec}(\mathbf{K}_k)^\top \end{bmatrix}^\top$ & $2n_z\!+\!n_x n_z$ \\
        Raw & $\begin{bmatrix} \boldsymbol{\nu}_k^\top, & \text{diag}(\mathbf{S}_k)^\top, & \text{vec}(\mathbf{K}_k)^\top \end{bmatrix}^\top$ & $2n_z\!+\!n_x n_z$ \\
        Whitened (no log) & $\begin{bmatrix} \mathbf{L}_k^{-1}\boldsymbol{\nu}_k^\top, & \text{vec}(\mathbf{K}_k)^\top \end{bmatrix}^\top$ & $n_z\!+\!n_x n_z$ \\
        NIS & $\begin{bmatrix} \boldsymbol{\nu}_k^\top \mathbf{S}_k^{-1}\boldsymbol{\nu}_k, & \text{vec}(\mathbf{K}_k)^\top \end{bmatrix}^\top$ & $1\!+\!n_x n_z$ \\ \hline
    \end{tabular}
\end{table}

\begin{table}[t!]
    \centering
    \caption{Performance of input variants. Mean $\pm$ std and median ARMSE across 100 training seeds.}
    \label{tab:ablation_input_perf}
    \small
    \begin{tabular}{lccc}
        \hline
        \textbf{Variant} & \textbf{Lorenz} & \multicolumn{2}{c}{\textbf{R{\"o}ssler (OOD)}} \\
                         & mean $\pm$ std [med] & mean $\pm$ std [med] & Div. \\ \hline
        Whitened (Proposed) & $0.661 \pm 0.648$ [$0.528$] & $0.872 \pm 1.299$ [$0.527$] & 3.57\% \\
        Raw & $0.783 \pm 0.799$ [$0.551$] & $1.364 \pm 1.760$ [$0.857$] & 4.97\% \\
        Whitened (no log) & $0.775 \pm 0.843$ [$0.530$] & $1.060 \pm 1.843$ [$\mathbf{0.476}$] & 5.20\% \\
        NIS & $0.598 \pm 0.501$ [$0.521$] & $\mathbf{0.766 \pm 0.959}$ [$0.551$] & $\mathbf{3.07\%}$ \\ \hline
    \end{tabular}
\end{table}

The \textbf{Raw} variant, which forfeits scale invariance, suffers the largest degradation in both mean ($1.364$) and median ($0.857$), confirming that scale invariance is essential for effective OOD generalization.

The \textbf{NIS} variant achieves the lowest mean ARMSE ($0.766$) and divergence rate ($3.07\%$) among all input representations, benefiting from a lower-dimensional optimization landscape. However, its median OOD error ($0.551$) exceeds that of the proposed representation ($0.527$), indicating that while the scalar statistic reduces worst-case failure frequency, it sacrifices typical-case accuracy. More fundamentally, collapsing the innovation vector into a scalar irreversibly destroys the directional error geometry required for dimension-wise covariance modulation. In systems where specific sensor axes degrade independently, a scalar statistic cannot attribute innovation anomalies to their causal channels, rendering NIS structurally incompatible with the vector-valued attenuation policy.

Among the multidimensional representations, the \textbf{Whitened (no log)} variant achieves the lowest R{\"o}ssler median ($0.476$) but suffers from the highest divergence rate ($5.20\%$) and an elevated mean error ($1.060$). This indicates susceptibility to poor initialization. The proposed \textbf{Whitened} representation integrates the logarithmic uncertainty magnitude, which acts as an implicit regularizer. This addition stabilizes the optimization landscape, improving the mean ARMSE to $0.872$ and reducing the divergence rate to $3.57\%$, ensuring robust convergence without sacrificing directional information.

\section{Experimental Validation: UAV State Estimation}\label{sec:uav_experiments}

While simulations confirm the algorithmic generalizability of the proposed framework, practical viability necessitates resilience against real-world unmodeled disturbances. We therefore validated the NDR-SHKF offline by processing recorded telemetry and ground-truth motion capture data from a physical UAV.

\subsection{Experimental Platform and State Formulation}
\begin{figure}[t!]
    \centering
    \includegraphics[width=1.0\columnwidth]{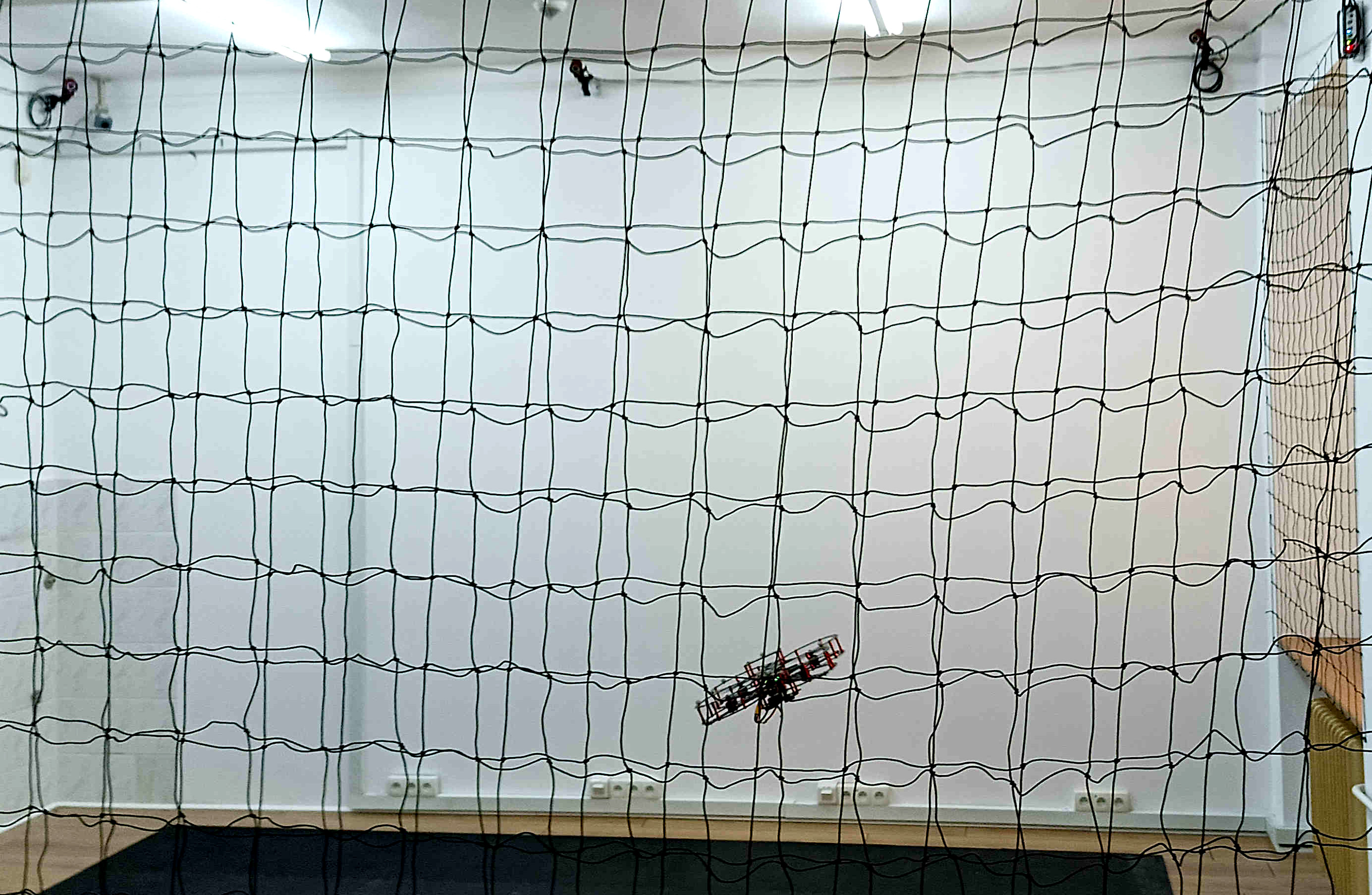}
    \caption{The Quanser QDrone experimental platform within the OptiTrack motion capture environment, which provides high-precision ground truth data.}
    \label{fig:platform}
\end{figure}

Experiments were conducted utilizing a Quanser QDrone quadrotor (Fig.~\ref{fig:platform}). The platform is equipped with an onboard Inertial Measurement Unit (IMU) operating at 1000 Hz, providing high-frequency specific force and angular rate measurements. Ground truth for performance evaluation was provided by an OptiTrack motion capture system yielding sub-millimeter accuracy~\cite{optitrack}. The physical and electrical parameters of the platform, directly utilized in the filter's dynamic models, are summarized in Table~\ref{tab:qdrone_parameters}~\cite{quanser2024qdrone}. 

The full 19-dimensional kinematic and dynamic state of the vehicle, denoted as $\mathbf{x}$, comprises its position in the Earth-fixed frame $\mathbf{r} =[r_x, r_y, r_z]^\top$, its velocity in the body frame $\mathbf{v} =[v_x, v_y, v_z]^\top$, its attitude expressed as a Hamiltonian quaternion $\mathbf{q} =[q_w, q_x, q_y, q_z]^\top$, its body angular rates $\boldsymbol{\omega} =[\omega_x, \omega_y, \omega_z]^\top$, and the biases of its onboard IMU sensors, $\mathbf{b}_a=[b_{ax}, b_{ay}, b_{az}]^\top$ and $\mathbf{b}_g=[b_{gx}, b_{gy}, b_{gz}]^\top$. The state vector is defined as:
\begin{equation}
    \mathbf{x} =[\mathbf{r}^\top, \mathbf{v}^\top, \mathbf{q}^\top, \boldsymbol{\omega}^\top, \mathbf{b}_a^\top, \mathbf{b}_g^\top]^\top.
\end{equation}

To evaluate the adaptability of the neural hyper-controller across distinct mathematical constraints, the framework was assessed under two disparate system modeling paradigms: an IMU-driven Kinematic formulation and a Control-driven Dynamic formulation.

\subsubsection{Model Formulation 1: IMU-Driven Kinematics}
In the kinematic formulation, the accelerometer ($\mathbf{a}^{\text{imu}} = [a^{\text{imu}}_{x}, a^{\text{imu}}_y, a^{\text{imu}}_z]^\top$) and gyroscope ($\boldsymbol{\omega}^{\text{imu}} = [\omega^{\text{imu}}_x, \omega^{\text{imu}}_y, \omega^{\text{imu}}_z]^\top$) measurements are employed as high-frequency control inputs $\mathbf{u}^{\text{imu}} = [\mathbf{a}^{{\text{imu}}^\top}, \boldsymbol{\omega}^{{\text{imu}}^\top}]^\top$ to the prediction step of the filter. The true specific force and angular rates are obtained by subtracting the estimated biases: $\tilde{\mathbf{a}} = \mathbf{a}^{\text{imu}} - \mathbf{b}_a$ and $\tilde{\boldsymbol{\omega}} = \boldsymbol{\omega}^{\text{imu}} - \mathbf{b}_g$. The continuous-time evolution of the states is given by:
\begin{align}
    \dot{\mathbf{r}} &= \mathbf{R}^{E}_{B}(\mathbf{q}) \mathbf{v}, \\[4pt]
    \dot{\mathbf{v}} &= \tilde{\mathbf{a}} + \mathbf{R}^{E}_{B}(\mathbf{q})^\top \mathbf{g} - \tilde{\boldsymbol{\omega}} \times \mathbf{v}, \\[4pt]
    \dot{\mathbf{q}} &= \frac{1}{2} \mathbf{q} \otimes \begin{bmatrix} 0 \\ \tilde{\boldsymbol{\omega}} \end{bmatrix}, \\[4pt]
    \dot{\boldsymbol{\omega}} &= \mathbf{0}, \quad \dot{\mathbf{b}}_a = \mathbf{0}, \quad \dot{\mathbf{b}}_g = \mathbf{0},
\end{align}
where $\mathbf{R}^{E}_{B}(\mathbf{q})$ represents the Direction Cosine Matrix (DCM) that defines the orientation of the drone, effectively transforming vectors from the Body-fixed frame ($\mathcal{F}_B$) to the Earth-fixed inertial frame ($\mathcal{F}_E$). Parameterized by the Hamiltonian quaternion $\mathbf{q} = [q_w, q_x, q_y, q_z]^\top$, this rotation matrix is explicitly formulated as:
\begin{equation}
    \mathbf{R}^{E}_{B}(\mathbf{q}) = \begin{bmatrix}
        1 - 2(q_y^2 + q_z^2) & 2(q_x q_y - q_w q_z) & 2(q_x q_z + q_w q_y) \\[4pt]
        2(q_x q_y + q_w q_z) & 1 - 2(q_x^2 + q_z^2) & 2(q_y q_z - q_w q_x) \\[4pt]
        2(q_x q_z - q_w q_y) & 2(q_y q_z + q_w q_x) & 1 - 2(q_x^2 + q_y^2)
    \end{bmatrix}.
\end{equation}
Conversely, its transpose $\mathbf{R}^{E}_{B}(\mathbf{q})^\top$ performs the inverse transformation from the Earth frame back to the Body frame. This inverse mapping is utilized to project the global gravity vector $\mathbf{g} = [0, 0, -g]^\top$ into the drone's $+Z$-up body axes during the velocity update. Furthermore, $\otimes$ denotes quaternion multiplication. 

While angular velocity $\boldsymbol{\omega}$ is mathematically redundant in this kinematics-driven formulation, it is retained with zero dynamics ($\dot{\boldsymbol{\omega}} = \mathbf{0}$) to maintain a constant 19-dimensional state vector ($n_x = 19$). This architectural consistency ensures the neural hyper-controller’s input and output dimensions remain invariant, enabling the same network to operate across both kinematic and control-driven models without structural modifications.

These state predictions are subsequently fused with global pose updates:
\begin{equation}
    \mathbf{z}_{k} = \begin{bmatrix}
    x_o,\ y_o,\ z_o,\ {\phi}_o,\ {\theta}_o,\ {\psi}_o
    \end{bmatrix}^\top.
\end{equation}

Although obtained from the high-precision OptiTrack system, these observations are intentionally corrupted with synthetic, heavy-tailed noise during both training and testing. This deliberate degradation provides a realistic surrogate for noisy positioning and heading telemetry, preventing the filter from developing an over-reliance on idealized sensor data. 

The observation model $h(\mathbf{x})$ defines the analytical mapping from the 19-dimensional state manifold to the 6-dimensional observation space ($\mathbb{R}^{19} \rightarrow \mathbb{R}^6$). Translational kinematics are mapped linearly by extracting the Cartesian position $\mathbf{r}$ directly from the state vector. Conversely, rotational kinematics require a nonlinear geometric transformation to project the Hamiltonian attitude $\mathbf{q}$, bounded on the unit hypersphere $S^3$, into the corresponding 3-2-1 Euler angle parameterization (roll $\phi$, pitch $\theta$, and yaw $\psi$). This projection yields the \textit{a priori} predicted measurement vector $\hat{\mathbf{z}}_{k|k-1}$, which is evaluated against the stochastic observation $\mathbf{z}_k$ to generate the innovation sequence $\boldsymbol{\nu}_k$ required for the EKF measurement update.

\subsubsection{Model Formulation 2: Control-Driven Dynamics}
In operational aerospace environments, external positioning signals such as GPS are vulnerable to attenuation, jamming, or complete outages \cite{yu2025electronic,matzke2025addressing}. To guarantee continuous, high-precision state estimation during such sensor-denied intervals, the estimation framework must be capable of transitioning to a strictly proprioceptive dead reckoning paradigm \cite{hashim2025advances}. Consequently, an alternative control-driven dynamic model is formulated~\cite{ireland2012development}. Instead of relying on external pose measurements to correct kinematic drift, this architecture employs a rigid-body representation that maps motor commands directly to state derivatives and predicted IMU readings.

The control input $\mathbf{u}^{\text{control}} = [F, \tau_x, \tau_y, \tau_z]^\top$ consists of the total thrust $F$ and the control torques $\boldsymbol{\tau} = [\tau_x, \tau_y, \tau_z]^\top$ generated by the rotors, which are mapped linearly from the motor percentage commands $\boldsymbol{\delta}=[\delta_1, \delta_2, \delta_3, \delta_4]^\top$ via the aerodynamic force and torque constants $K_f$ and $K_t$:
\begin{equation}
\begin{bmatrix}
    F \\
    \tau_{x} \\
    \tau_{y} \\
    \tau_{z}
\end{bmatrix}
=
\begin{bmatrix}
    K_f & K_f & K_f & K_f \\[6pt]
    -K_f \frac{L_{roll}}{2} & -K_f \frac{L_{roll}}{2} & K_f \frac{L_{roll}}{2} & K_f \frac{L_{roll}}{2} \\[6pt]
    K_f \frac{L_{pitch}}{2} & -K_f \frac{L_{pitch}}{2} & K_f \frac{L_{pitch}}{2} & -K_f \frac{L_{pitch}}{2} \\[6pt]
    K_t & -K_t & -K_t & K_t
\end{bmatrix}
\begin{bmatrix}
    \delta_1 \\
    \delta_2 \\
    \delta_3 \\
    \delta_4
\end{bmatrix}
\end{equation}
The continuous-time evolution of the states is given by:
\begin{align}
    \dot{\mathbf{r}} &= \mathbf{R}^{E}_{B}(\mathbf{q}) \mathbf{v}, \\[4pt]
    \dot{\mathbf{v}} &= \mathbf{f}_b + \mathbf{R}^{E}_{B}(\mathbf{q})^\top \mathbf{g} - \boldsymbol{\omega} \times \mathbf{v}, \\[4pt]
    \dot{\mathbf{q}} &= \frac{1}{2} \mathbf{q} \otimes \begin{bmatrix} 0 \\ \boldsymbol{\omega} \end{bmatrix}, \\[4pt]
    \dot{\boldsymbol{\omega}} &= \mathbf{J}^{-1} (\boldsymbol{\tau} - \boldsymbol{\omega} \times \mathbf{J} \boldsymbol{\omega}), \\[4pt]
    \dot{\mathbf{b}}_a &= \mathbf{0}, \quad \dot{\mathbf{b}}_g = \mathbf{0},
\end{align}
where $\mathbf{J}$ is the UAV's inertia matrix and $\mathbf{f}_b$ represents the net specific force acting on the body, accounting for both thrust and linear aerodynamic drag parameterized by the diagonal matrix $\mathbf{D}$:
\begin{equation}
    \mathbf{f}_b = \frac{1}{M} \begin{bmatrix} 0 \\ 0 \\ F \end{bmatrix} - \mathbf{D} \mathbf{v}.
\end{equation}
Because the QDrone is operated in a controlled indoor motion-capture environment with minimal ambient wind, translational aerodynamic drag is considered negligible~\cite{silveria2025lie}. However, to maintain strict numerical stability within the Jacobians, it is modeled using a minimal uniform drag coefficient, $\mathbf{D} = \text{diag}(0.001, 0.001, 0.001) \text{ N}\cdot\text{s/m}$. Unlike the kinematic model, the measurement function $h(\mathbf{x}, \mathbf{u})$ in this dynamic formulation projects the internal state into the proprioceptive sensor space, generating predicted 6-DoF IMU readings by combining the physical states with the estimated sensor biases:
\begin{equation}
    h(\mathbf{x}, \mathbf{u}) = \hat{\mathbf{z}}_{k|k-1} = \begin{bmatrix}
    (\mathbf{f}_b + \mathbf{b}_a)^\top, \
    (\boldsymbol{\omega} + \mathbf{b}_g)^\top
    \end{bmatrix}^\top.
\end{equation}

Consequently, the true stochastic observation $\mathbf{z}_k \in \mathbb{R}^6$ supplied to the filter during a sensor-denied interval is composed strictly of the raw accelerometer and gyroscope signals:
\begin{equation}
    \mathbf{z}_{k} = \begin{bmatrix}
    \mathbf{a}^{\text{imu}\top}, \
    \boldsymbol{\omega}^{\text{imu}\top}
    \end{bmatrix}^\top.
\end{equation}

\begin{table}[t!]
    \centering
    \caption{Key physical and aerodynamic parameters of the QDrone~\cite{quanser2024qdrone}.}
    \label{tab:qdrone_parameters}
    \begin{tabular}{llc}
        \hline
        \textbf{Symbol} & \textbf{Description} & \textbf{Value} \\
        \hline
        \multicolumn{3}{c}{\textbf{Mass and Inertia}} \\
        \hline
        $M$ & Total mass & 1.121 kg \\
        $g$ & Gravitational acceleration & 9.81 m/s$^2$ \\
        $J_{xx}$ & Roll Moment of Inertia & $1.00 \times 10^{-2}$ kg$\cdot$m$^2$ \\
        $J_{yy}$ & Pitch Moment of Inertia & $8.20 \times 10^{-3}$ kg$\cdot$m$^2$ \\
        $J_{zz}$ & Yaw Moment of Inertia & $1.48 \times 10^{-2}$ kg$\cdot$m$^2$ \\
        \hline
        \multicolumn{3}{c}{\textbf{Dimensions and Motor Constants}} \\
        \hline
        $L_{roll}$ & Roll motor-to-motor distance & 0.2136 m \\
        $L_{pitch}$ & Pitch motor-to-motor distance & 0.1758 m \\
        $u_{hover}$ & Nominal hover motor command & 53.8\% \\
        $K_f$ & Motor thrust coefficient & 5.11 N \\
        $K_t$ & Motor torque coefficient & 0.0487 N$\cdot$m \\ \hline
    \end{tabular}
\end{table}

\subsection{Training Methodology and Domain Randomization}
The NDR-SHKF was optimized utilizing an aggregated dataset comprising five distinct empirical flight trajectories recorded from the QDrone platform, totaling approximately 6 minutes and 28 seconds of flight data (yielding roughly 388,000 discrete samples at 1000 Hz). To guarantee that the recurrent policy encountered a diverse distribution of dynamic state transitions, these training flights consisted of unstructured, dynamic movements, including step responses, and varying acceleration profiles. 

To prevent the neural architecture from overfitting to contiguous flight logs, training sequences were extracted as random overlapping windows sampled proportionally to the size of each respective dataset. This dynamic sampling strategy ensures that any segment of the flight data has a non-zero probability of being selected at any point during training, effectively eliminating the potential bias introduced by fixed contiguous chunking. To stabilize the BPTT dynamics and prevent gradient explosion within the deeply unrolled EKF formulation, a curriculum learning strategy was implemented for the sequence length $T$. Training commenced with short sequences to allow the meta-policy to learn fundamental noise scales and maintain numerical stability, gradually expanding to longer horizons to capture slow-accumulating IMU bias drift and steady-state covariance behavior. Specifically, the sequence length was scheduled as follows: $T=20$ for the first 300 epochs, $T=50$ for epochs 300 to 599, $T=100$ for epochs 600 to 899, $T=200$ for epochs 900 to 1999, and finally $T=300$ for epochs 2000 and beyond. Optimization was executed over a total of 2200 epochs, with each epoch evaluating 5 batches.

The neural hyper-controller is trained \textit{exclusively} under Model Formulation 1 (IMU-Driven Kinematics). During this offline optimization phase, the filter utilizes the raw high-frequency IMU data directly as control inputs to propagate the 19-dimensional state ($n_x=19$), while updating against synthetic, artificially degraded OptiTrack measurements acting as 6-dimensional observations ($n_z=6$). The network never encounters Model Formulation 2 (Control-Driven Dynamics) during training; its ability to subsequently transition to a purely proprioceptive dead reckoning paradigm during a loss of external positioning references relies entirely on the out-of-distribution generalizability of the learned memory attenuation policy.

To accommodate the expanded state and measurement dimensions of the UAV platform, the input vector $\mathbf{y}_k$ is scaled to $\mathbb{R}^{126}$. The neural architecture retains the core network capacity established in Section~\ref{sec:sim_experiments}, specifically the 3-layer GRU stack with 32 hidden units and the bifurcated context routing. However, the policy head is widened to two 32-neuron layers to output the corresponding 25-dimensional adaptation vector $\mathbf{d}_k$. The model was trained end-to-end via BPTT utilizing the Adam optimizer (with a multi-stage piecewise decay schedule), a batch size of 128, and a global gradient clipping threshold of 0.1 enforced to maintain stability throughout the recursive EKF updates.

To stabilize the gradient flow and prevent anomalous predictions from disproportionately skewing the optimization landscape during early epochs, the primary state estimation objective was bifurcated into translational and rotational components. The translational error was penalized via a robust Huber loss ($\mathcal{L}_{\text{pos}}$)~\cite{huber1992robust}. Let $\mathbf{e}_k = \hat{\mathbf{r}}_k - \mathbf{r}_k$ denote the position error vector at time step $k$, with individual spatial components $e_{k,i}$. The sequence-averaged translational loss is evaluated element-wise as:
\begin{equation}
    \mathcal{L}_{\text{pos}} = \frac{1}{T} \sum_{k=1}^T \left( \frac{1}{n} \sum_{i=1}^{n} \text{Huber}_{\delta_H} (e_{k,i}) \right),
\end{equation}
where the scalar Huber loss is defined as:
\begin{equation}
    \text{Huber}_{\delta_H}(e) = 
    \begin{cases} 
        \frac{1}{2} e^2 & \text{if } |e| \le \delta_H, \\
        \delta_H \left( |e| - \frac{1}{2} \delta_H \right) & \text{otherwise.}
    \end{cases}
\end{equation}
Here, $\delta_H = 5.0$ dictates the threshold governing the transition from a quadratic to a linear error penalty, effectively bounding the magnitude of the gradients induced by large spatial outliers.

The rotational tracking error was evaluated using a quaternion cosine distance metric. To account for the double-cover topology of unit quaternions, where $\mathbf{q}$ and $-\mathbf{q}$ represent the same physical orientation in 3D space, the formulation utilizes the absolute value of the inner product. This ensures that the loss penalizes the shortest-path angular misalignment between the predicted attitude $\hat{\mathbf{q}}_k$ and the ground truth $\mathbf{q}_k$, bounding the penalty between 0 (perfect alignment) and 1 (maximum deviation):
\begin{equation}
    \mathcal{L}_{\text{att}} = \frac{1}{T} \sum_{k=1}^T \left( 1 - |\langle \hat{\mathbf{q}}_k, \mathbf{q}_k \rangle| \right).
\end{equation}

Synthesizing the kinematic tracking objectives with the shallow-context auxiliary reconstruction loss ($\mathcal{L}_{\text{aux}}$) to explicitly prevent feature collapse, the composite multi-task loss function evaluated over each trajectory sequence is formulated as:
\begin{equation}
    \mathcal{L}_{\text{total}} = \mathcal{L}_{\text{pos}} + \lambda_{\text{att}} \mathcal{L}_{\text{att}} + \lambda_{\text{aux}} \mathcal{L}_{\text{aux}},
\end{equation}
where the weighting hyperparameters were empirically established as $\lambda_{\text{att}} = 10.0$ and $\lambda_{\text{aux}} = 0.1$.

\subsubsection{Stochastic Domain Randomization}
To address the simulation-to-reality gap and improve the performance of the neural hyper-controller in the presence of unmodeled dynamics, continuous domain randomization was incorporated during training. This approach promotes the development of robust estimation policies by preventing the network from relying on idealized telemetry.

The external pose observations $\mathbf{z}_k$ and high-frequency IMU inputs $\mathbf{u}_k$ (comprising specific force and body angular rates) were processed through physically motivated degradation models. Let $\tilde{\mathbf{z}}_k$ and $\tilde{\mathbf{u}}_k$ denote the stochastically perturbed measurement and input vectors, respectively. The degradation sequence is defined as follows:
\begin{align}
    \tilde{\mathbf{z}}_k &= \mathbf{z}_k + \mathbf{v}_k^{\text{meas}}, \\
    \tilde{\mathbf{u}}_k &= \left( \mathbf{u}_k + \mathbf{v}_k^{\text{inp}} \right) \odot (\mathbf{1} + \mathbf{s}) + \mathbf{b}^{\text{init}} + \mathbf{b}_k^{\text{walk}} + \mathbf{v}_k^{\text{vib}}.
\end{align}

Each term represents a common electromechanical defect or environmental anomaly, detailed in Table~\ref{tab:domain_randomization}. External measurements are intentionally corrupted by heavy-tailed outliers to serve as a surrogate for sensor degradation. Proprioceptive inputs are degraded by baseline thermo-mechanical noise punctuated by high-amplitude transient faults, static multiplicative distortions from cross-axis misalignments, and turn-on bias instability driven by thermal variations. Additionally, low-frequency non-stationary random walks capture long-term operational instability, while harmonic vibrations model the unmodeled high-frequency structural noise mechanically coupled from the rotor dynamics. To ensure varied noise exposure and prevent deterministic memorization, the magnitude of each noise parameter is dynamically sampled per batch element from a uniform distribution, $\mathcal{U}(\text{ratio} \times \text{max}, \text{max})$, where the ratio dictates the minimum bound. Furthermore, while positional measurements are subjected to measurement outliers, attitude measurements are restricted to Gaussian noise to accurately reflect realistic orientation constraints.

\begin{table}[t!]
    \centering
    \caption{Stochastic Domain Randomization Models}
    \label{tab:domain_randomization}
    \small
    \resizebox{\columnwidth}{!}{
    \renewcommand{\arraystretch}{1.4}
        \begin{tabular}{p{0.28\columnwidth} p{0.50\columnwidth} p{0.18\columnwidth}}\hline
        \textbf{Mechanism} & \textbf{Mathematical Formulation} & \textbf{Parameters} \\ \hline
        \multicolumn{3}{c}{\textbf{External Pose Measurements}} \\ \hline
        Measurement\newline Outlier \cite{roth2013student} & $\mathbf{v}_k^{\text{meas}} \sim (1-\epsilon)\mathcal{N}(\mathbf{0}, \sigma_{\text{meas}}^2\mathbf{I}) + \epsilon\mathcal{N}(\mathbf{0}, \sigma_{\text{out}}^2\mathbf{I})$ & $\sigma_{\text{meas}} = 1.0$ \newline $\sigma_{\text{out}} = 5.0$ \newline $\epsilon = 0.05$ \\ \hline
        \multicolumn{3}{c}{\textbf{Proprioceptive IMU Inputs}} \\ \hline
        Transient\newline Sensor Faults \cite{el2008analysis} & $\mathbf{v}_k^{\text{inp}} \sim (1-\epsilon)\mathcal{N}(\mathbf{0}, \sigma_{\text{inp}}^2\mathbf{I}) + \epsilon\mathcal{N}(\mathbf{0}, \sigma_{\text{out}}^2\mathbf{I})$ & $\sigma_{\text{inp}} = 0.1$ \newline $\sigma_{\text{out}} = 5.0$ \newline $\epsilon = 0.01$ \\
        Scale Factor\newline \& Misalignment\newline \cite{titterton2004strapdown, woodman2007introduction} & $\mathbf{s} \sim \mathcal{N}(\mathbf{0}, \sigma_{\text{sf}}^2\mathbf{I})$ & $\sigma_{\text{sf}} = 10^{-3}$ \\
        Turn-On Bias\newline Instability \cite{titterton2004strapdown} & $\mathbf{b}^{\text{init}} \sim \mathcal{N}(\mathbf{0}, \sigma_{\text{bias}}^2\mathbf{I})$ & $\sigma_{\text{bias}} = 10^{-3}$ \\
        Bias Rate\newline Random Walk \cite{el2008analysis} & $\mathbf{b}_k^{\text{walk}} = \sum_{i=1}^k \mathbf{w}_i, \quad \mathbf{w}_i \sim \mathcal{N}(\mathbf{0}, \sigma_{\text{walk}}^2\mathbf{I})$ & $\sigma_{\text{walk}} = 10^{-5}$ \\
        Rotor-Induced\newline Vibration \cite{leishman2014quadrotors, brossard2020ai} & $\mathbf{v}_k^{\text{vib}} = \mathbf{n}_k \odot (1 + A \sin^2(\omega k \Delta t))$ \newline $\mathbf{n}_k \sim \mathcal{N}(\mathbf{0}, \sigma_{\text{vib}}^2\mathbf{I})$ & $A=0.1$ \newline $\omega = 628.0$ \newline $\sigma_{\text{vib}} = 1.0$ \\ \hline
        \end{tabular}
    }
\end{table}

By structuring the algebraic sequence of these degradations, where base noise is multiplicatively scaled prior to the accumulation of thermal drift and structural vibrations, the estimator is forced to learn robust, out-of-distribution representations. 

\subsection{Benchmarking: Noise Amplification and Sensor Outages}

To evaluate cross-domain generalizability, we conducted a benchmark on three distinct hold-out circular trajectories (each 10 seconds) using recorded flight telemetry. These trajectories impose sustained centripetal acceleration and rotational dynamics, ensuring the estimator generalizes to the underlying flight physics rather than overfitting to the training paths. The evaluation protocol simulates three scenarios of progressive sensor degradation using these recorded logs:

\begin{enumerate}
    \item \textbf{Baseline Scenario:} Evaluates tracking performance under standard operating conditions, characterized by nominal measurement noise and ambient mechanical vibrations, supplemented by the training-phase measurement noise injection.
    \item \textbf{Transient Disturbance Scenario:} Simulates a localized, 2-second increase in measurement uncertainty (scaling the position and attitude measurement noise variance by a factor of 2.0) to simulate intermittent signal interference. The filter must detect this shift and attenuate noisy updates to prevent divergence.
    \item \textbf{Sensor-Denied Scenario:} Introduces a complete, 2-second loss of external positioning data during the 10-second circular maneuver. The estimators were transitioned from an observation-based kinematic model to a proprioceptive dynamic model, relying exclusively on internal IMU integration and motor control commands during the outage.
\end{enumerate}

We compared the NDR-SHKF ($N=3$ GRU layers) against a standard EKF and two classical Sage-Husa variants ($b=0.995$ and $0.999$) to contrast fixed parameter tuning against dynamic adaptation. To ensure a rigorous comparison, all estimators were identically initialized with the same state $\mathbf{x}_0$, error covariance $\mathbf{P}_0$, and nominal noise matrices $\mathbf{Q}$ and $\mathbf{R}$.

We restrict the flight data benchmark to classical and adaptive filters (EKF, SHKF). The Variational Bayesian AKF (VBAKF) exhibited systematic divergence during preliminary trials due to the tight coupling between its Inverse-Wishart hyperprior and the strong nonlinearities of the 19-dimensional state. Learning-based baselines (ENN-AKF, L-EKF, KalmanNet) were excluded from the physical flight benchmark because they fundamentally lack the recursive structural safeguards required to bound state estimates under unmodeled 19-DoF dynamics. Applying unbounded neural gain synthesis to physical hardware operating outside its training distribution risks rapid, unconstrained divergence. Modifying these architectures to operate safely in real-world 19-DoF flight constitutes a separate research endeavor. Evaluating them in this context without such architectural overhauls would likely fail and provides no scientifically meaningful comparison against bounded covariance methods.

The initial state vector $\mathbf{x}_0$ was seeded using the first available ground-truth observation to populate the position and attitude (converted from Euler angles to a Hamiltonian quaternion), while all dynamic derivatives and IMU biases (velocity, angular velocity, $\mathbf{b}_a$, $\mathbf{b}_g$) were initialized to zero. The initial \textit{a posteriori} error covariance was defined as a block-diagonal matrix:
\begin{equation}
    \mathbf{P}_0 = \text{diag}\big(10^{-1} \mathbf{I}_{3}, \ 1.0 \mathbf{I}_{3}, \ 10^{-1} \mathbf{I}_{4}, \ 10^{-1} \mathbf{I}_{3}, \ 10^{-4} \mathbf{I}_{3}, \ 10^{-3} \mathbf{I}_{3}\big),
\end{equation}
corresponding to the uncertainty in position, velocity, quaternion, angular velocity, accelerometer bias, and gyroscope bias, respectively.

The nominal process noise covariance $\mathbf{Q}$ was structurally defined to reflect realistic, varying confidence levels across the distinct physical state dimensions:
\begin{equation}
    \mathbf{Q} = \text{diag}\big(10^{-2} \mathbf{I}_{3}, \ 10^{-1} \mathbf{I}_{3}, \ 10^{-2} \mathbf{I}_{4}, \ 10^{-1} \mathbf{I}_{3}, \ 10^{-5} \mathbf{I}_{3}, \ 10^{-5} \mathbf{I}_{3}\big),
\end{equation}

Similarly, the measurement noise covariance was configured homogeneously across all 6 spatial and rotational observations:

\begin{equation}
    \mathbf{R} = 5 \times 10^{-2} \mathbf{I}_{6}.
\end{equation}

\subsection{Benchmark Results and Analysis}
\begin{figure*}[ht!]
    \centering
    \includegraphics[width=1.0\linewidth]{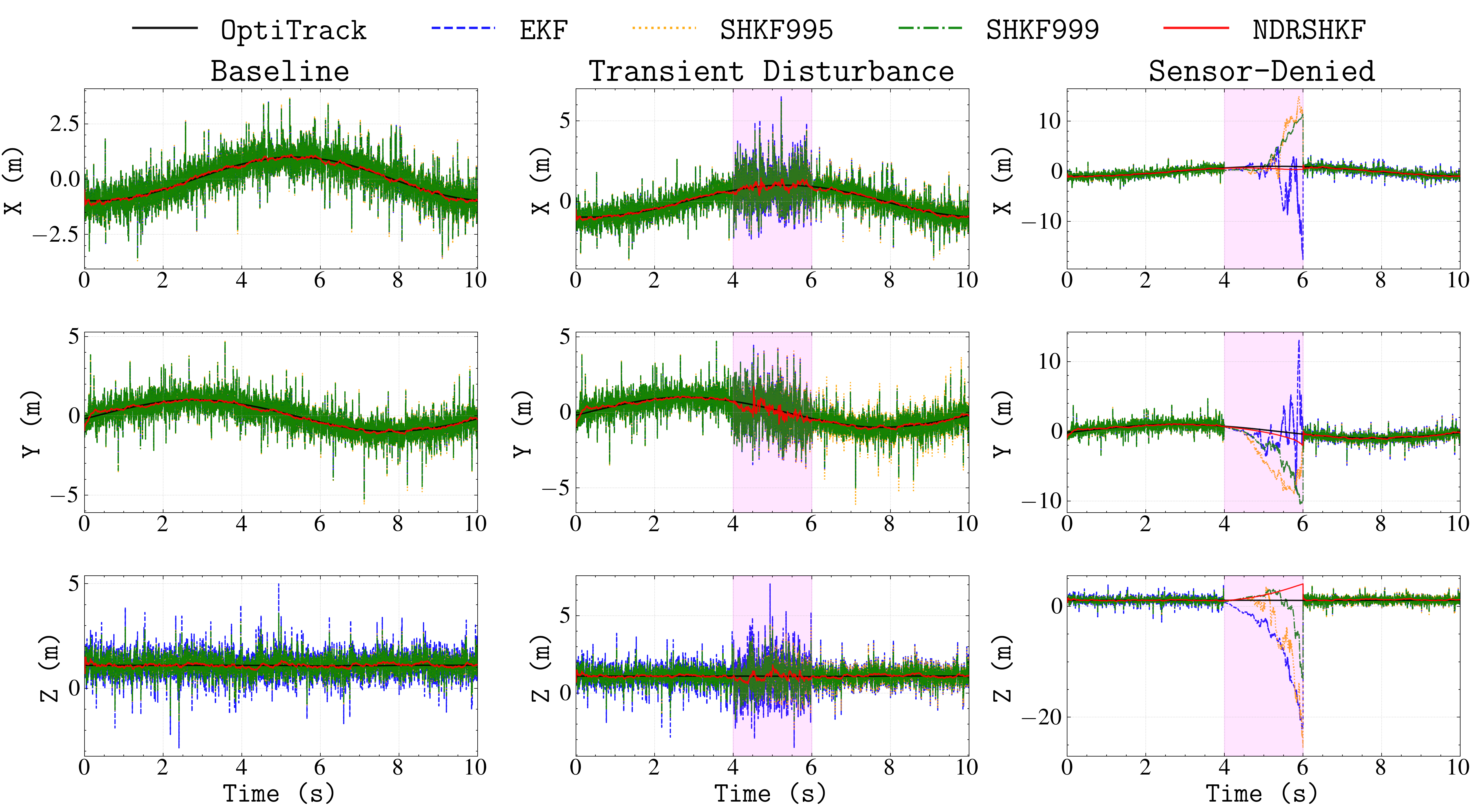}
    \caption{UAV Position tracking (X, Y, Z) during the benchmark flight. The magenta shaded regions indicate the 2-second intervals where the transient disturbance and the sensor outage are applied in their respective scenarios.}
    \label{fig:uav_xyz}
\end{figure*}

\begin{figure*}[ht!]
    \centering
    \includegraphics[width=1.0\linewidth]{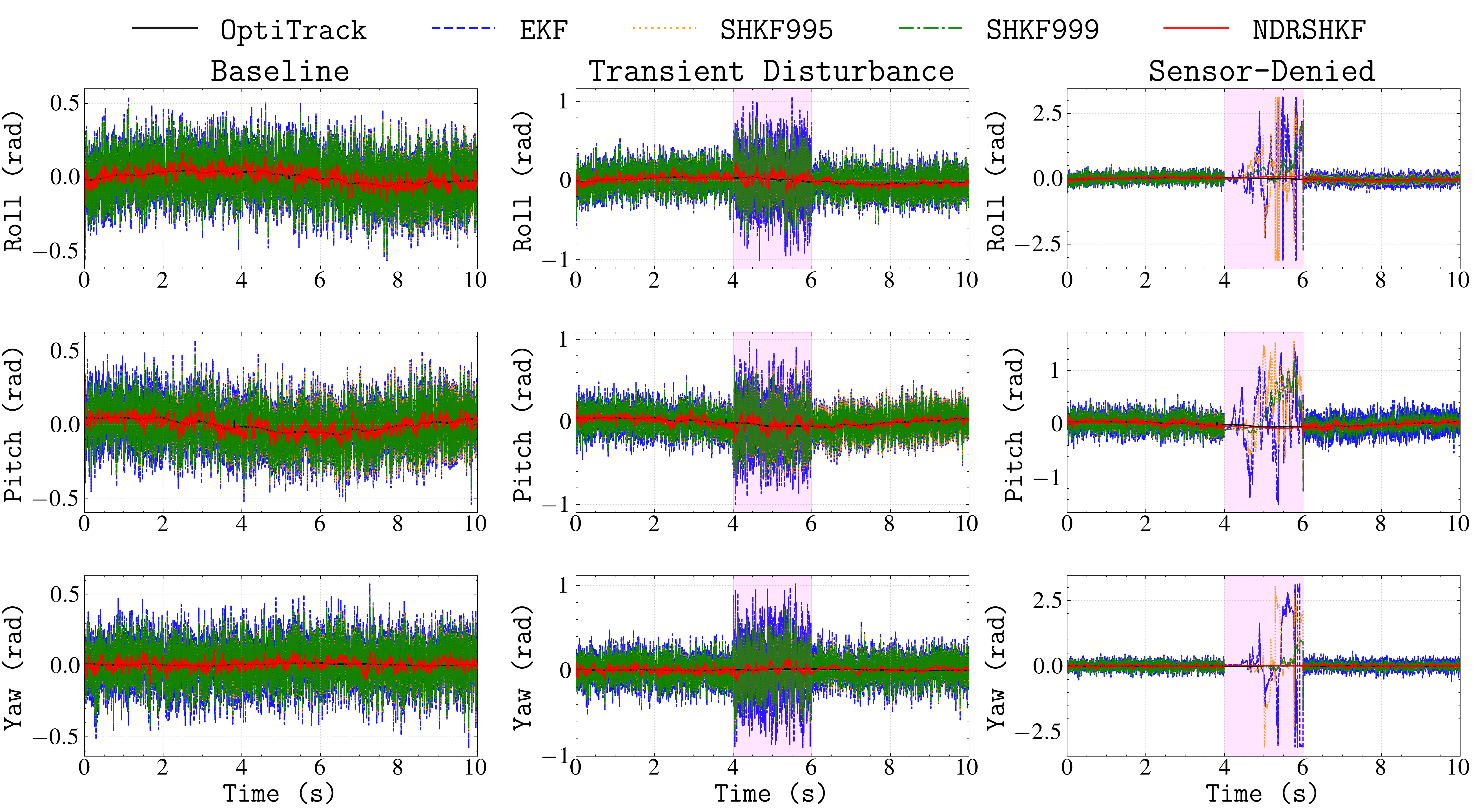}
    \caption{UAV Attitude tracking (Roll, Pitch, Yaw). The magenta shaded regions denote the active disturbance and blackout intervals. The NDR-SHKF maintains orientation during the simulated sensor blackout better than the traditional EKF and SHKF variants.}
    \label{fig:uav_attitude}
\end{figure*}

\begin{figure}[ht!]
    \centering
    \includegraphics[width=1.0\linewidth]{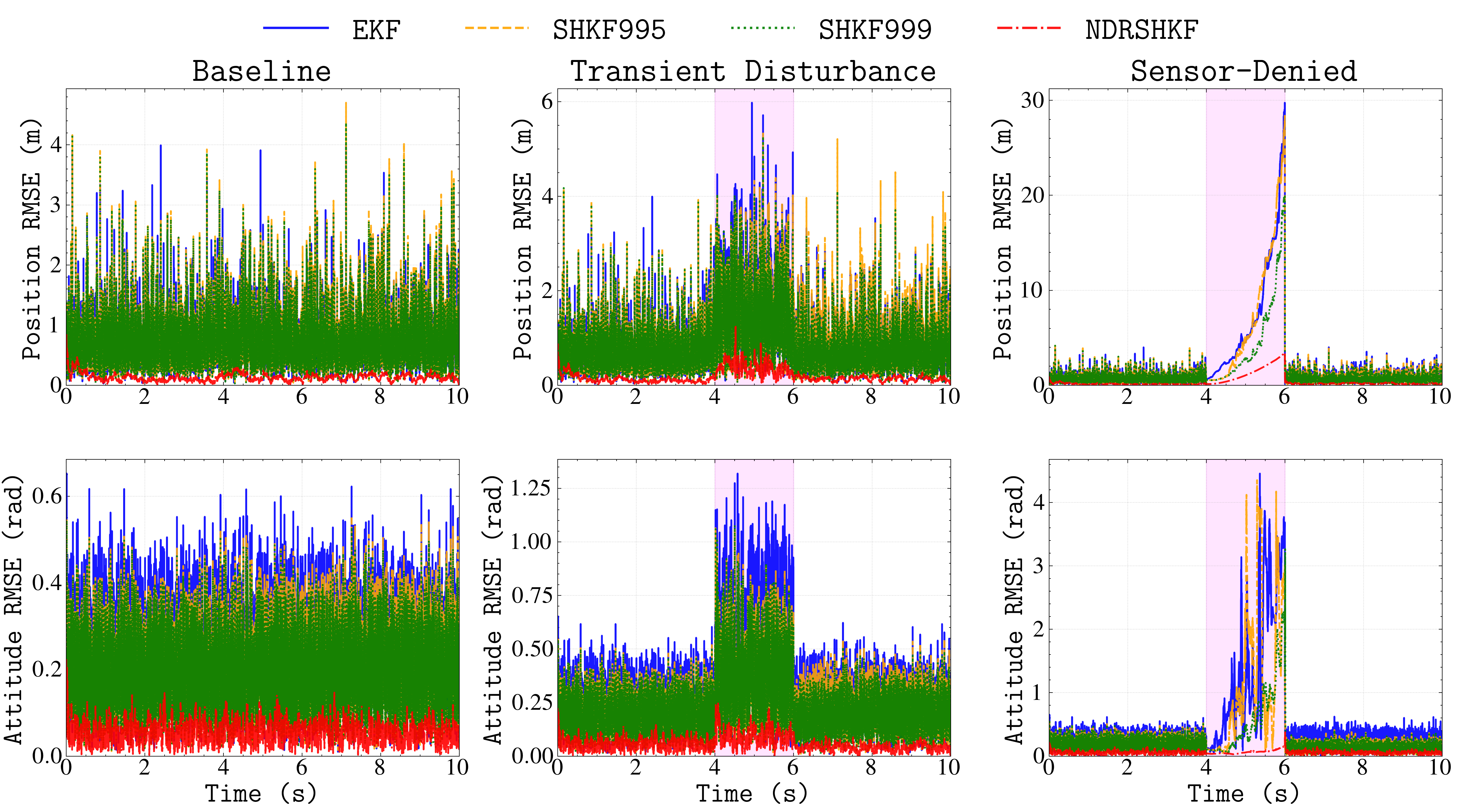}
    \caption{UAV Position and Attitude RMSE over time. The magenta shaded regions indicate the injection of the transient disturbance and sensor outage. Classical estimators suffer significant divergence without external corrections, whereas the NDR-SHKF dynamically modulates its confidence bounds to arrest compounding drift during the sensor outage.}
    \label{fig:uav_pos_and_att_rmse}
\end{figure}

Table~\ref{tab:uav_benchmark} summarizes the position and attitude RMSE across the three evaluation modes. Trajectories and temporal error magnitudes are depicted in Fig.~\ref{fig:uav_xyz}, Fig.~\ref{fig:uav_attitude}, and Fig.~\ref{fig:uav_pos_and_att_rmse}.

\begin{table}[t!]
\centering
\caption{UAV Flight Benchmark Results (mean $\pm$ std over 3 trajectories). The \textit{Transient Disturbance} scenario applies a $2\times$ variance distortion to external measurements, while the \textit{Sensor-Denied} scenario forces a complete external measurement blackout requiring transition to proprioceptive dead reckoning.}
\label{tab:uav_benchmark}
\renewcommand{\arraystretch}{1.2}
\begin{tabular}{lcc}
\hline
\textbf{Method} & \textbf{Pos. RMSE (m)} & \textbf{Att. RMSE (rad)} \\
\hline
\multicolumn{3}{c}{\textbf{Baseline}} \\
\hline
EKF      & $0.537 \pm 0.069$ & $0.147 \pm 0.010$ \\
SHKF995  & $0.552 \pm 0.083$ & $0.133 \pm 0.004$ \\
SHKF999  & $0.529 \pm 0.070$ & $0.128 \pm 0.005$ \\
NDR-SHKF & $\mathbf{0.091 \pm 0.012}$ & $\mathbf{0.035 \pm 0.001}$ \\
\hline
\multicolumn{3}{c}{\textbf{Transient Disturbance}} \\
\hline
EKF      & $0.678 \pm 0.044$ & $0.195 \pm 0.005$ \\
SHKF995  & $0.668 \pm 0.061$ & $0.160 \pm 0.002$ \\
SHKF999  & $0.635 \pm 0.051$ & $0.157 \pm 0.004$ \\
NDR-SHKF & $\mathbf{0.127 \pm 0.009}$ & $\mathbf{0.039 \pm 0.004}$ \\
\hline
\multicolumn{3}{c}{\textbf{Sensor-Denied}} \\
\hline
EKF      & $3.235 \pm 1.382$ & $0.470 \pm 0.069$ \\
SHKF995  & $3.451 \pm 0.598$ & $0.431 \pm 0.024$ \\
SHKF999  & $2.393 \pm 0.711$ & $0.291 \pm 0.045$ \\
NDR-SHKF & $\mathbf{0.463 \pm 0.032}$ & $\mathbf{0.058 \pm 0.030}$ \\
\hline
\end{tabular}
\end{table}

During the \textit{Baseline} scenario, all filters tracked the state successfully. However, the NDR-SHKF exhibited superior noise attenuation, yielding a position RMSE of $0.091 \pm 0.012$ m compared to $0.529 \pm 0.070$ m for the best classical baseline (SHKF999), an over 80\% reduction in error.

The limitations of fixed adaptation parameters were evident during the \textit{Transient Disturbance} scenario. The classical EKF assimilated degraded external measurements without adjustment, increasing position error to $0.678 \pm 0.044$ m. The SHKF995 over-amplified instantaneous sensor noise due to its aggressive adaptation rate, degrading attitude estimates. Conversely, the NDR-SHKF identified the error expansion within the whitened innovation space and selectively attenuated the update, maintaining a position RMSE of $0.127 \pm 0.009$ m. This is approximately one-fifth that of the SHKF999 ($0.635 \pm 0.051$ m), with strictly separated standard deviation bounds.

In the \textit{Sensor-Denied} scenario, the EKF exhibited rapid position drift ($3.235 \pm 1.382$ m) because it lacks corrective measurement updates to arrest the compounding IMU integration error. The SHKF995 showed high cross-trajectory variance ($3.451 \pm 0.598$ m), demonstrating sensitivity to initial conditions, while the SHKF999, despite being the strongest classical baseline across all scenarios ($2.393 \pm 0.711$ m), could not arrest the compounding drift as effectively as the learned policy. The NDR-SHKF modulated its covariance to account for the telemetry loss, minimizing drift and maintaining a position RMSE of $0.463 \pm 0.032$ m. This represents a 7-fold reduction compared to the EKF and an over $5\times$ improvement over the SHKF999.

The hierarchical recurrent context of the NDR-SHKF enabled the adaptation policy to correctly identify the sustained bias in the innovation sequence caused by the loss of external pose references. By scaling the process noise covariance and modifying the influence of the measurement residuals, the filter transitioned to proprioceptive dead reckoning without accumulating unbounded drift. Upon restoration of the external positioning telemetry, the attenuation vector reconverged to nominal values, re-establishing tracking accuracy.

Beyond state estimation robustness, the offline flight-log profiling directly validates the complexity analysis of Section~\ref{subsec:complexity}. At the UAV operating point ($n_x = 19$, $n_z = 6$, $N = 3$, $d_h = 32$), the dominant EKF term ($n_x^3 \approx 6859$) and the recurrent neural term ($N d_h^2 \approx 3072$) are of comparable magnitude, placing the architectural overhead within the same complexity class as the base filter. Benchmarking across three independent flight segments confirms that the NDR-SHKF incurs only a 42\% latency increase over an EKF ($53.2\,\mu\text{s}$ versus $37.4\,\mu\text{s}$ per step), satisfying the $1\,\text{ms}$ control-loop of the platform. The dense linear-algebraic operations that dominate both the EKF Jacobian evaluations and the GRU inference share identical computational primitives, ensuring direct compatibility with existing vectorized embedded inference frameworks without requiring specialized kernels~\cite{lai2018cmsis,stone2026real}.

\subsection{Cross-Dynamics Transfer Benchmark}
\label{subsec:cross_dynamics_transfer}

To evaluate the structural generalization of the learned memory attenuation policy, we perform a transfer experiment. The NDR-SHKF model trained exclusively on the 19-dimensional physical UAV flight dataset (Section~\ref{sec:uav_experiments}) is evaluated directly on the topologically distinct, 3-dimensional Lorenz and R{\"o}ssler chaotic attractors from Section~\ref{sec:sim_experiments}.

Because the UAV-trained policy expects a 126-dimensional input vector derived from an $n_x^{\text{uav}}=19$, $n_z^{\text{uav}}=6$ state-space, evaluating it on an $n_x^{\text{sim}}=3$, $n_z^{\text{sim}}=2$ domain requires embedding the low-dimensional inputs into the high-dimensional representation. We implement a semantic zero-fill mapping utilizing discrete positional offsets for the state ($x_{\text{off}}$) and measurement ($z_{\text{off}}$) channels. Assuming standard 0-indexed, right-exclusive array slicing, the mapped input vector $\mathbf{y}_{\text{uav}} \in \mathbb{R}^{126}$ is populated as follows:
\begin{align}
    \mathbf{y}_{\text{uav}}[z_{\text{off}} : z_{\text{off}} + n_z^{\text{sim}}] &= \tilde{\boldsymbol{\nu}}_{\text{sim}}, \\
    \mathbf{y}_{\text{uav}}[n_z^{\text{uav}} + z_{\text{off}} : n_z^{\text{uav}} + z_{\text{off}} + n_z^{\text{sim}}] &= \mathbf{l}_{\text{sim}}, \\
    \mathbf{K}_{\text{uav}}[x_{\text{off}} : x_{\text{off}} + n_x^{\text{sim}},\; z_{\text{off}} : z_{\text{off}} + n_z^{\text{sim}}] &= \mathbf{K}_{\text{sim}}, \\
    \mathbf{y}_{\text{uav}}[2n_z^{\text{uav}} :] &= \text{vec}(\mathbf{K}_{\text{uav}}),
\end{align}
where $\mathbf{K}_{\text{uav}} \in \mathbb{R}^{19 \times 6}$ is an otherwise zero matrix, and $\tilde{\boldsymbol{\nu}}_{\text{sim}}$ and $\mathbf{l}_{\text{sim}}$ denote the whitened innovation and logarithmic uncertainty vectors of the simulation domain, respectively. The output adaptation factors are correspondingly extracted strictly from the active mapped channels:
\begin{equation}
    \mathbf{d}_k^Q = \mathbf{d}_k[x_{\text{off}} : x_{\text{off}} + n_x^{\text{sim}}], \quad \mathbf{d}_k^R = \mathbf{d}_k[n_x^{\text{uav}} + z_{\text{off}} : n_x^{\text{uav}} + z_{\text{off}} + n_z^{\text{sim}}].
\end{equation}

Table~\ref{tab:cross_dyn} reports the results of 10,000 Monte Carlo runs evaluating two structural embedding strategies. The first strategy, a coupled linear shift, proportionately scales the state and measurement index offsets ($x_{\text{off}} = 4 z_{\text{off}}$). The second strategy, an attitude-isolated decoupling, anchors the abstract simulation measurements strictly to the UAV's physical orientation channels ($z_{\text{off}} = 3$) while independently translating the state mapping across the rows of the Kalman gain matrix.

\begin{table}[t!]
    \centering
    \caption{Cross-Dynamics Transfer Benchmark. ARMSE across 10,000 Monte Carlo runs evaluating a UAV-trained policy ($n_x=19$, $n_z=6$) on the Lorenz and R{\"o}ssler attractors ($n_x=3$, $n_z=2$). Baseline divergence of $\sim$2.2\% on the R{\"o}ssler system stems from inherent ground-truth instability, consistent with Section~\ref{sec:sim_experiments}.}
    \label{tab:cross_dyn}
    \small
    \renewcommand{\arraystretch}{1.2}
    \begin{tabular}{ccccc}
        \hline
        $z_{\text{off}}$ & $x_{\text{off}}$ & \textbf{Lorenz} & \multicolumn{2}{c}{\textbf{R{\"o}ssler}} \\ 
         & & mean $\pm$ std & mean $\pm$ std & Div. (\%) \\ \hline
        \multicolumn{5}{c}{\textbf{Coupled Linear Shift} ($x_{\text{off}} = 4 z_{\text{off}}$)} \\ \hline
        0 & 0 & $0.545 \pm 0.129$ & $0.678 \pm 0.672$ & 2.24\% \\
        1 & 4 & $0.558 \pm 0.165$ & $0.864 \pm 0.963$ & 2.23\% \\
        2 & 8 & $0.574 \pm 0.127$ & $0.605 \pm 0.612$ & 2.23\% \\
        3 & 12 & $0.569 \pm 0.129$ & $0.849 \pm 0.831$ & 2.25\% \\
        4 & 16 & $0.565 \pm 0.173$ & $0.756 \pm 0.713$ & 2.24\% \\ \hline
        \multicolumn{5}{c}{\textbf{Attitude-Isolated Decoupling} ($z_{\text{off}} = 3$)} \\ \hline
        3 & 0 & $0.580 \pm 0.212$ & $0.515 \pm 0.351$ & 2.23\% \\
        3 & 8 & $0.589 \pm 0.183$ & $0.581 \pm 0.409$ & 2.23\% \\
        3 & 16 & $0.566 \pm 0.149$ & $0.799 \pm 0.812$ & 2.23\% \\ \hline
    \end{tabular}
\end{table}

The transfer benchmark yields ARMSE values directly competitive with models explicitly trained on the chaotic domains (Table~\ref{tab:results}). The architecture maintains estimation stability despite the zero-padding of the Kalman gain array and arbitrary semantic shifting of the active channels. Structural ablation analyses, including artificial disturbance injection and analytical Jacobian evaluation, confirm this robustness is not the result of channel-agnostic thresholding. Instead, the network learns physical cross-channel coupling (e.g., modulating translational process noise during attitude measurement disturbances) but extracts this multivariate dependency almost entirely from the whitened innovation vector ($\tilde{\boldsymbol{\nu}}_k$). Because the Cholesky factor $\mathbf{L}_k^{-1}$ projects the raw residuals into a mixed, scale-invariant Mahalanobis space, cross-channel error geometries are encoded prior to the neural forward pass. This structural pre-mixing allows the policy to generalize efficiently across diverse mathematical bounds without strict reliance on the explicit positional structure of the raw Kalman gain.

\section{Conclusion}\label{sec:conclusions}
This paper demonstrated that replacing the static forgetting factor of the Sage-Husa filter with a learned, vector-valued attenuation policy yields improved out-of-distribution generalization compared to classical heuristics. By operating on whitened innovation sequences, the NDR-SHKF successfully maps multi-scale temporal dynamics to dimension-specific adaptation rates. Offline evaluations on UAV flight datasets confirmed the framework bridges the transition to proprioceptive dead reckoning during sensor-denied intervals, a condition where classical adaptive filters suffer rapid error accumulation or divergence. The current architecture is restricted to diagonal covariance matrices. While enforcing diagonal structures with explicit algebraic safeguards guarantees numerical stability, this constraint mathematically precludes the estimation of cross-channel noise correlations. Future work will address this by expanding the neural policy to predict dense covariance matrices via symmetric positive-definite parameterizations and by transitioning the algorithm from offline dataset evaluation to closed-loop hardware deployment for real-time benchmarking.

\bibliographystyle{elsarticle-num}
\bibliography{paper}

\end{document}